\documentclass[12pt]{oxarticle}
\pdfoutput=1

\usepackage{graphicx, float, array, xspace, amscd, amsmath, amsthm, amssymb, latexsym}
\usepackage[bbgreekl]{mathbbol}
\usepackage{amsfonts,bbm}

\DeclareSymbolFontAlphabet{\mathbb}{AMSb}
\DeclareSymbolFontAlphabet{\mathbbl}{bbold}

\usepackage[sort&compress, comma, square, numbers]{natbib}

\usepackage{color}
\definecolor{MyDarkBlue}{rgb}{0.15,0.25,0.45}
\usepackage[linktocpage=true]{hyperref}
\hypersetup{colorlinks=true, citecolor=blue, linkcolor=blue, urlcolor=blue, pdfauthor={}, pdftitle={},breaklinks=true}

\let\SS=\S 

\def\Psib{{\overline \Psi}}

\def\ON{{||\O||}}
\newcommand{\adot}{{\dot{a}}}
\newcommand{\bdot}{{\dot{b}}}

\newcommand{\ddc}{\text{d}^{\text c}}
\newcommand{\ol}{\overline}
\def\starb{\,{\overline{ \star}\,}}
\newcommand{\id}{{\mathbbm{1}}}
\newcommand{\eb}{{\overline{\eta}}}
\newcommand{\zeb}{{\overline{\zeta}}}
\newcommand{\xb}{{\overline{\xi}}}
\renewcommand{\sb}{{\overline{\sigma}}}

\newcommand{\rb}{{\overline{ r}}}

\newcommand{\Bb}{{\overline{B}}}

\newcommand{\Xib}{{\overline{\Xi}}}

\newcommand{\veb}{{{\overline{\varepsilon}}\,}}

\newcommand{\Mb}{{\overline{M}}}
\newcommand{\Nb}{{\overline{N}}}

\newcommand{\ad}{{\rm ad }}

\newcommand{\Ob}{{\overline{ \Omega}}}

\newcommand{\cCb}{{\overline{\mathcal{C}}}}
\newcommand{\cDb}{{\overline{\mathcal{D}}}}

\newcommand{\w}{{\,\wedge\,}}
\newcommand{\wt}{\widetilde}

\newcommand{\Le}{{\mathfrak{e}}}
\newcommand{\Lh}{{\mathfrak{h}}}
\newcommand{\Lg}{{\mathfrak{g}}}

\newcommand{\su}{{\mathfrak{su}}}
\newcommand{\so}{{\mathfrak{so}}}

\newcommand{\Dslash}{\ensuremath \raisebox{0.025cm}{\slash}\hspace{-0.32cm} D}

\newcommand{\half}{\frac{1}{2}}

\def\rep#1{{{\boldsymbol{#1}}}}
\def\brep#1{{{\overline{\boldsymbol{#1}}}}}

\def\CS{{\text{CS}}}

\newcommand{\ab}{{\overline\alpha}}
\newcommand{\bb}{{\overline\beta}}
\newcommand{\gb}{{\overline\gamma}}

\renewcommand{\a}{\alpha}
\renewcommand{\b}{\beta}
\newcommand{\g}{\gamma}\newcommand{\G}{\Gamma}
\renewcommand{\d}{\delta}\newcommand{\D}{\Delta}
\newcommand{\e}{\epsilon}\newcommand{\ve}{\varepsilon}
\newcommand{\z}{\zeta}

\renewcommand{\th}{\theta}\newcommand{\Th}{\Theta}

\renewcommand{\k}{\kappa}
\renewcommand{\l}{\lambda}\renewcommand{\L}{\Lambda}
\newcommand{\m}{\mu}
\newcommand{\n}{\nu}
\newcommand{\x}{\xi}

\renewcommand{\r}{\rho}
\newcommand{\s}{\sigma}\renewcommand{\S}{\Sigma}
\renewcommand{\t}{\tau}

\newcommand{\Ph}{\Phi}
\newcommand{\ch}{\chi}

\renewcommand{\o}{\omega}\renewcommand{\O}{\Omega}

\DeclareFontFamily{OT1}{pzc}{}
\DeclareFontShape{OT1}{pzc}{m}{it}{<-> s * [1.200] pzcmi7t}{}
\DeclareMathAlphabet{\mathpzc}{OT1}{pzc}{m}{it}

\newcommand{\cA}{\mathcal{A}}
\newcommand{\ccB}{\mathpzc B}
\newcommand{\cC}{\mathcal{C}}
\newcommand{\cD}{\mathcal{D}}\newcommand{\ccD}{\mathpzc D}
\newcommand{\cE}{\mathcal{E}}\newcommand{\ccE}{\mathpzc E}
\newcommand{\cF}{\mathcal{F}}

\newcommand{\cH}{\mathcal{H}}\newcommand{\ccH}{\mathpzc H}

\newcommand{\cK}{\mathcal{K}}\newcommand{\ccK}{\mathpzc K}
\newcommand{\cL}{\mathcal{L}}
\newcommand{\cM}{\mathcal{M}}\newcommand{\ccM}{\mathpzc M}
\newcommand{\cN}{\mathcal{N}}
\newcommand{\cO}{\mathcal{O}}

\newcommand{\cR}{\mathcal{R}}

\newcommand{\ccT}{\mathpzc T}

\newcommand{\ccW}{\mathpzc W}
\newcommand{\ccX}{\mathpzc X}
\newcommand{\cY}{\mathcal{Y}}\newcommand{\ccY}{\mathpzc Y}\newcommand{\cYb}{{\overline \cY}\,}
\newcommand{\cZ}{\mathcal{Z}}\newcommand{\ccZ}{\mathpzc Z}

\newcommand{\IC}{\mathbbl{C}}

\newcommand{\IF}{\mathbbl{F}}

\newcommand{\IR}{\mathbbl{R}}

\font\eightrm=cmr8 at 8pt
\font\csc=cmcsc10

\newcommand{\beq}{\begin{equation}}
\newcommand{\eeq}{\end{equation}}
\newcommand{\beqnn}{\begin{equation*}}
\newcommand{\eeqnn}{\end{equation*}}
\newcommand{\bea}{\begin{eqnarray}}
\newcommand{\eea}{\end{eqnarray}}
\newcommand{\bean}{\begin{eqnarray*}}
\newcommand{\eean}{\end{eqnarray*}}

\newcommand{\sref}[1]{\SS\ref{#1}}

\newcommand{\pd}[2]{\frac{\partial #1}{\partial #2}}

\newcommand{\norm}[1]{\left\| #1\right\|}

\newcommand{\ii}{\text{i}}

\newcommand{\place}[3]{\vbox to0pt{\kern-\parskip\kern-7pt
                             \kern-#2truein\hbox{\kern#1truein #3}
                             \vss}\nointerlineskip}
\newcommand{\smallfrac}[2]{\frac{\scriptstyle #1}{\scriptstyle #2}}

\DeclareFontFamily{U}{wncy}{}
\DeclareFontShape{U}{wncy}{m}{n}{<->wncyr10}{}
\DeclareSymbolFont{mcy}{U}{wncy}{m}{n}
\DeclareMathSymbol{\sha}{\mathord}{mcy}{"58}

\newcommand{\eu}[1]{{\mathfrak #1}}

\newcommand{\capt}[3]{\parbox{#1}{\renewcommand{\baselinestretch}{1.0}
                                                           \caption{\label{#2}\small\it #3}}}

\newcommand{\del}{\partial}
\newcommand{\delb}{\overline{\partial}}

\newcommand{\jb}{{\bar\jmath}}

\newcommand{\nb}{{\bar\n}}
\newcommand{\mb}{{\bar\m}}

\newcommand{\delbA}{\delb_{{}\hskip-2.8pt\cA}}

\newcommand{\A}{\cA}

\newcommand{\dd}{\text{d}}

\newcommand{\cym}{Calabi-Yau manifold\xspace}

\newcommand{\K}{K\"ahler\xspace}

\newcommand{\vol}{\dd^6x g^\frac12}

\newcommand{\tr}{\,\text{Tr}\,}

\newcommand{\tb}{{\overline{\tau}}}

\newcommand{\ap}{{\a^{\backprime}\,}}
\renewcommand{\sb}{{\overline{\sigma}}}
\renewcommand{\rb}{{\overline{\rho}}}

\makeatletter
\g@addto@macro\bfseries{\boldmath}
\makeatother

\renewcommand{\baselinestretch}{1.1}
\numberwithin{equation}{section}
\begin{document}
\pagestyle{empty}

\ifproofmode\underline{\underline{\Large Working notes. Not for circulation.}}\else{}\fi

\begin{center}
\null\vskip0.2in
{\Huge On the Effective Field Theory \\

of Heterotic Vacua\\[0.5in]}
{\csc 
Jock McOrist$\,^{1}$
\\[0.5in]}
 %
%
{\it Department of Mathematics\hphantom{$^2$}\\
University of Surrey\\
Guildford, GU2 7XH, UK\\
}
\footnotetext[1]{{\tt j.mcorist@surrey.ac.uk}}
\vskip2cm

{\bf Abstract\\}
\end{center}
The effective field theory of heterotic vacua that realise $\IR^{3,1}$  preserving $\cN{=}1$ supersymmetry are studied. The vacua in question admit  large radius limits taking the form $\IR^{3,1}\times \ccX$, with $\ccX$ a smooth three-fold with vanishing first Chern class and a stable holomorphic gauge bundle $\ccE$.  In a previous paper we calculated the kinetic terms for moduli, deducing the moduli metric and \K potential. In this paper, we compute the remaining couplings in the effective field theory, correct to first order in $\ap$. In particular, we compute the contribution of the matter sector to the \K potential, derive the Yukawa couplings and other quadratic fermionic couplings. From this we write down a  \K potential $\ccK$ and superpotential $\ccW$.

%
\vspace*{-5pt}

\newpage
\renewcommand{\baselinestretch}{1.4}\tableofcontents
\newpage
\setcounter{page}{1}
\pagestyle{plain}


\section{Introduction}
We are interested in heterotic vacua that realise $\cN{=}1$ supersymmetric field theories in $\IR^{3,1}$. At large radius, these take form $\IR^{3,1}\times \ccX$ where $\ccX$ is a compact smooth complex three-fold with vanishing first Chern class. We study the $E_8{\times} E_8$ heterotic string, and so there is a holomorphic vector bundle $\ccE$ with a structure group $\ccH\subset E_8\times E_8$ and a $d{=}4$ spacetime gauge symmetry given by the commutant  $\eu{G} = [E_8{\times} E_8, \ccH]$. The bundle $\ccE$ has a connection $A$, with field strength $F$ satisfying the hermitian Yang-Mills equation. The field strength $F$ is related to a gauge-invariant three-form $H$ and the curvature of $\ccX$ through anomaly cancellation.  The triple $(\ccX, \ccE, H)$ forms a heterotic structure, and the moduli space of these structures is described by what we call heterotic geometry. In this paper, we compute the contribution of fields charged under the spacetime gauge group $\eu{G}$ to the heterotic geometry.

The challenge in studying heterotic vacua is the complicated relationship between $H$, the field strength $F$ and the geometry of $\ccX$. Supersymmetry relates the complex structure $J$ and Hermitian form $\o$ of $\ccX$ to the gauge invariant three-form $H$:
\beq
 H ~=~ \dd^c \o~, \qquad  \dd^c\o ~=~ \half J_{m_1}{}^{n_1}J_{m_2}{}^{n_2} J_{m_3}{}^{n_3} (\del_{n_1}\o_{n_2n_3})\, \dd x^{m_1} \dd x^{m_2} \dd x^{m_3}~.
\label{eq:SusyRelation0}\eeq
where $x^m$ are real coordinates on $\ccX$. Green--Schwarz anomaly cancellation gives  a modified Bianchi identity for $H$
\beq
\dd H =- \frac{\ap}{4} \left( \tr F^2 - \tr R^2\, \right)~,
\label{eq:Anomaly0}\eeq
where in the second of these equations $R$ is the curvature two-form computed with respect to a appropriate connection with torsion proportional to $H$. This means the tangent bundle $\ccT_\ccX$ has torsion if $H$ is non-zero. Unless one is considering the standard embedding --- in which $\ccE$ is identified with $\ccT_\ccX$ the tangent bundle to $\ccX$  --- the right hand side of \eqref{eq:Anomaly0} is non-zero even when $\ccX$ is a \cym at large radius. This means that $H$ is generically non-vanishing, though subleading in $\ap$, and so even for large radius heterotic vacua $\ccX$ is non-\K. Torsion is inescapable.  

The effective field theory of the light fields for these vacua are described by a Lagrangian with $\cN=1$ supersymmetry,  whose bosonic sector is of the form
\beq
\cL ~=~ \frac{1}{2\k_4^2} \sqrt{-G_4} \left(\cR_4 -\frac{1}{4}\tr |F_\Lg|^2 - 2G_{A {\overline B}} \widehat\cD_e \Phi^A \widehat\cD_e \Phi^{\overline B} - V(\Phi,\bar\Phi) + \cdots  \right)~.\label{eq:EFT1}
\eeq
Here $\k_4$ is the four-dimensional Newton constant,  $\cR_4$ the four-dimensional Ricci-scalar,  $F_\Lg$ is the spacetime gauge field strength, the $\Phi^A$ range over the scalar fields of the field theory and their kinetic term comes with a metric $G_{A\Bb}$. The fields $\Phi^A$ may  be charged under $\Lg$, the algebra of the gauge group $\eu{G}$, with an appropriate covariant derivative  $\widehat\cD_e$.  Finally $V(\Phi,\bar\Phi)$ is the bosonic potential for the scalars. 

When $\ccE\cong \ccT_\ccX$ the moduli space of the heterotic theory reduces to that of a \cym, and is described by special geometry. The unbroken gauge group in spacetime is $E_6$, and the charged matter content consists of fields charged in the $\rep{27}$ and $\brep{27}$ representations. The Yukawa couplings were calculated in supergravity in for example \cite{Strominger:1985it,Strominger:1985ks}. The effective field theory of this compactification was described in a beautiful paper  \cite{Dixon:1989fj}, in which  relations between the \K potential and superpotential were computed using string scattering amplitudes, $(2,2)$ supersymmetry and Ward identities. The \K and superpotential were shown to be related to each other and in fact, were both determined in terms of a pair of holomorphic functions.  These are known as the special geometry relations. For a  review of special geometry in the language of this paper see \cite{Candelas:1989bb}.  A key question is how these relations generalise to other choices of bundle $\ccE$.  

We work towards answering this question by computing the effective field theory couplings correct to first order in $\ap$. In a previous paper  \cite{Candelas:2016usb} we commenced a study of heterotic geometry using $\ap$-corrected supergravity. This is complementary to a series of papers  \cite{delaOssa:2014cia, delaOssa:2014msa,delaOssa:2015maa,Anderson:2014xha,Garcia-Fernandez:2015hja} who identified the parameter space with certain cohomology groups.  In the context of effective field theory \eqref{eq:EFT1}, one of the results of \cite{Candelas:2016usb}  was to calculate the contribution of the bosonic moduli fields to the metric $G_{A\Bb}$. In this paper, we compute the contribution of the matter sector to the metric $G_{A\Bb}$, and the Yukawa couplings, correct to order $\ap$. We describe an ansatz for the superpotential and \K potential for effective field theory:
\beq
\begin{split}
 \ccK ~&=~   - \log\left( \frac{4}{3} \int \o^3\right)  -\log\left( \ii \!\int\! \O\, \Ob \right) + G_{\x\eb}  \tr C^{\x} C^{\eb}  + G_{\r\tb}\tr  D^{\t } D^{\rb }, \cr
\ccW &= -\ii \sqrt{2} e^{-\ii\phi} \int \Omega\Big( H - \dd^c \o\Big)~.
\end{split}\label{eq:SuperKahler}
\eeq 
The superpotential is normalised by comparing with the Yukawa couplings computed in the dimensional reduction using the conventions of Wess--Bagger \cite{WessBagger}.

The moduli have a metric
\beq
\begin{split}\label{eq:ModuliMetricIntro}
 \dd s^2 ~&=~  2G_{\a\bb} \,\dd y^\a  \otimes  \dd y^\bb~, \\[6pt]
G_{\a\bb} ~&=~  \frac{1}{4V} \int \D_\a{}^\m \star \D_\bb{}^\n \,\, g_{\m\nb} + \frac{1}{4V} \int \cZ_\a \star \cZ_\bb \,+ \\[3pt]
&\quad + \frac{ \ap}{4V}\int \tr \Big( D_\a A \star D_\bb A \Big) - \frac{ \ap}{4V}\int \tr \Big(D_\a \Th\,\star\, D_\bb \Th^\dag\Big)~,
\end{split}
\eeq
where $\cZ_\a = \ccB_\a + \ii \del_\a \o$ is the $\ap$-corrected, gauge invariant generalisation of the complexified Kahler form $\d B + \ii \d \o$, the $\chi_\a$ form a basis of closed $(2,1)$-forms, and the the last line is the Kobayashi metric, extended to the entire parameter space, including deformations of the spin connection on $\ccT_\ccX$. The metric expressed this way is an inner product of tensors corresponding to  complex structure $\D_\a$, hermitian moduli $\ccZ_\a$, and bundle moduli $D_\a A$.  The role of the spin connection $D_\a \th$  is presumably determined in terms of the other moduli as they do not correspond to independent physical fields. The tensors depend on parameters holomorphically through
 \beq
 \label{eq:ModuliHolomorphy}
\D_\a{}^\nb ~=~ 0, \qquad \cZ_\ab ~=~ \ccB_\ab + \ii \del_\ab \o ~=~ 0~, \quad \cD_\ab A^{0,1} ~=~ 0~, \qquad\cD_\ab \th^{0,1} ~=~ 0~.
 \eeq
de la Ossa and Svanes \cite{delaOssa:2014cia} show that there exists a choice of basis for the parameters in which each of the tensors in the metric are in an appropriate cohomology, \footnote{I would like to thank Xenia de la Ossa for explaining this choice of basis to me. }
%
%
hence, the moduli space metric \eqref{eq:ModuliMetricIntro} is the natural inner product (Weil--Peterson) on cohomology classes. 

The matter fields are $C^\x$ and $D^\t$  and appear in the \K potential trivially, as they do in special geometry. The matter metric is the Weil-Petersson inner product of corresponding cohomology elements 
\beq
\begin{split}
G_{\t\sb} ~&=~  
 \frac{\ap}{4V} \int_\ccX    \psi_\t  \,\star \,\psi_\sb~,\qquad G_{\x\eb} ~=~ 
\frac{\ap}{4V} \int_\ccX  \phi_\x  \, \star\,\phi_\eb\, ~, \\
\end{split}\label{eq:MatterMetricIntro}
\eeq
where $\phi_\x, \psi_\r$ are $(0,1)$-forms valued in a sum over representations of the structure group $\cH$. 

In some sense it was remarkable that one was able to find a compact closed expression for the \K potential for the moduli metric. This was not a priori obvious, especially given the non-linear PDEs relating parameters in the anomolous Bianchi identity and supersymmetry relations \eqref{eq:SusyRelation0}-\eqref{eq:Anomaly0}. Indeed, it turned out that the \K potential for the moduli in  \eqref{eq:SuperKahler} is of the same in form as that of  special geometry, except where one has replaced the \K form by the hermitian form $\o$. At first sight this is confusing as the only fields appearing in the \K potential are $\o$ and $\O$. Nonetheless, the \K potential still depends on bundle moduli in precisely the right way through a non-trivial  analysis of the supersymmetry and anomaly conditions. The hermitian form $\o$ contains, hidden within, information about both the bundle and hermitian moduli. 
\footnote{ It is important to note that the derivation here and in \cite{Candelas:2016usb}, no assumption is made about expanding around the standard embedding.  $\cE$ is not related to the tangent bundle}

 The metric \eqref{eq:ModuliMetricIntro}   is compatible with the result in \cite{Anguelova:2010ed}, who studied the $\ap$ and $\ap^2$ corrections to the moduli space metric in the particular case where the hermitian part of the metric varies, while the remaining fields are fixed: $(\del_a \o)^{1,1} \ne 0$, $\ccB_a = \D_a = D_a A = 0$. In the general all fields vary with parameters and the metric is non-zero already at $\cO(\ap)$.  
 

The analysis in \cite{Candelas:2016usb} focussed primarily on D-terms relevant to moduli. In this paper, we compute the remaining D-terms, including the metric terms for the bosonic matter fields charged under $\Lg$. We also compute the F-terms up to cubic order in fields, exploiting the formalism constructed in \cite{Candelas:2016usb}. The primary utility of this is to derive an expression for the Yukawa couplings in a manifestly covariant fashion. Together with the metrics discussed above, one is now finally able to compute properly normalised Yukawa couplings, relevant to to any serious particle phenomenology. The F-terms are protected in $\ap$-perturbation theory, and so the only possible $\ap$-corrections are due to worldsheet instantons. 

The fields neutral under $\Lg$, the singlet fields, also do not have any mass or cubic Yukawa couplings. In fact, all singlet couplings necessarily vanish. They correspond to moduli which are necessarily free parameters and so the singlets need to have unconstrained vacuum expectation values. If there were a non-zero singlet coupling at some order in the field expansion e.g. $\rep{1}^n$, or in a $\Le_6$ theory $(\rep{27}\cdot\brep{27})^{326}\cdot \rep{1}^{101}$ then some parameter $y^\a$ would have its value fixed, a contradiction on it being a free parameter.%
\footnote{An important open question is, when are singlet couplings are generated by worldsheet instantons? At least for vacua derived from linear sigma models, there are arguments that suggest that after summing over all worldsheet instantons all the singlet couplings vanish  \cite{Beasley:2003fx,Silverstein:1995re}. Here we assume the vacua is well-defined with a large radius limit, and so all singlet couplings vanish. }

The superpotential $\ccW$ in \eqref{eq:SuperKahler}  is an ansatz designed to replicate these couplings. Its functional form can be partly argued by symmetry. There is a complex line bundle over the moduli space in which the holomorphic volume form on $\ccX$, denoted $\O$, transforms with a gauge symmetry $\O \to \mu \O$ where $\mu \in \IC^*$. The superpotential is also a section of this line bundle, and transforms in the same way $\ccW \to \mu \ccW$. Hence, $\ccW$ has an integrand  proportional to $\O$. To make the integrand a nice top-form we need to wedge it with a gauge invariant three-form. The three-form needs to contain a dependence on the matter fields, and this can only occur through the ten-dimensional $H$ field. The other natural gauge invariant three-forms, that are not defined in a given complex structure are $\dd \o$ and $\dd^c\o$. $\ccW$ is also required not to give rise to any singlet couplings. So all derivatives of $\ccW$ with respect to parameters must vanish. The  combination $H - \dd^c \o$ manifestly satisfies this request.  Derivatives with respect to matter fields of $\ccW$ do not vanish. As these are charged in $\Lg$, the only non-zero contributions come from $H$. This allows us to fix the normalisation of $\ccW$ by comparing with the dimensional reduction calculation of the Yukawa couplings. Finally, $\ccW$ must be a holomorphic function of chiral fields, which is straightforward to check. It is convenient that the single expression for the superpotential captures both the matter and moduli couplings, and fact seemingly not realised before. 

 A complementary perspective on $\ccW$ was studied by \cite{delaOssa:2015maa}. In that paper, one starts with an $\su(3)$--structure manifold $\ccX$, posit the existence of $\ccW$, and use it as a device to  reproduce the conditions needed for the heterotic vacuum to be supersymmetry. This builds on earlier work in the literature, for example \cite{Becker:2003yv, Gurrieri:2004dt,LopesCardoso:2003dvb}. The superpotential ansatzed in those papers are of a different form to that described here, and the cubic and higher order singlet couplings nor Yukawa couplings were not consistently computed. We choose to work with the expression above as it manifestly replicates the vanishing of all singlet couplings.

The layout of this paper is the following. In \sref{s2} we review the necessary background to study heterotic vacua, reviewing the results of \cite{Candelas:2016usb}. In  \sref{s3}, we dimensionally reduce the Yang-Mills sector to obtain a metric on the matter fields. In  \sref{s4}, the reduction is applied to the gaugino to get the quadratic fermionic couplings, including the Yukawa couplings. In \sref{s5}, we summarise the results. In \sref{s6} we  show how these couplings are represented in the language of a \K potential $\ccK$ and superpotential $\ccW$.

\subsection{Tables of notation}
\begin{table}[H]
\def\bigstr{\vrule height22pt depth12pt width0pt}
\def\medstr{\vrule height19pt depth10pt width0pt}
\def\smallstr{\vrule height17pt depth9pt width0pt}
\begin{center}
\begin{tabular}{| >{$\hskip3pt} l <{\hskip3pt$} | >{\hskip3pt}l<{\hskip3pt} | 
>{\hskip3pt}l<{\hskip3pt} |}
\hline
\medstr \hfil \text{Quantity} &\hfil Definition \hfil &\hfil Comment\hfil\\ 
\hline\hline
\smallstr \Lg       &    $d{=}4$ spacetime gauge algebra      &group is $\eu{G}$      \\
\hline
\smallstr \Lh              &          structure algebra of $\ccE$  & group is $\ccH$         \\
\hline
\smallstr \rep{r}               &     representation of $\Lh$       &   $\dim \rep{r}= r$     \\
\hline
\smallstr \rep{R}             &     representation of $\Lg$         &   $\dim \rep{R} = R$      \\
\hline
\smallstr \Phi            &  $d=10$ gauge field  in $(\rep{r},\brep{R})$ of $\Lh\oplus\Lg$       &   $\phi = \Phi^{0,1}$, ~~$\Phi = \phi - \psi^\dag$     \\
\hline
\smallstr \Psi          &  $d=10$ gauge field in $(\brep{r},\rep{R})$ of $\Lh\oplus\Lg$           &   $\psi = \Psi^{0,1}$, ~~ $\Psi = \psi - \phi^\dag$   \\
\hline
\smallstr \phi_{\x}          &  basis for $H^1(\ccX,\ccE_{\rep{r}} )$         & valued in $\rep{r}$ of $\Lh$  \\
\hline
\smallstr \psi_{\r}         &  basis for $H^1(\ccX,\ccE_{\brep{r}} )$         &  valued in $\brep{r}$ of $\Lh$ \\
\hline
\smallstr C^{\x}, D^{\r}, Y^\a          &  $d{=}4$ bosons in the $\brep{R}$, $\rep{R}$, $\rep{1}$ of $\Lg$ & $\x,\t,\a$ label harmonic bases\\
&  (e.g.  $\brep{27}$, $\rep{27}$, $\rep{1}$ of $\Le_6$)         &  \\
\hline
\smallstr \cC^\x, \cD^\r  , \cY^\a         &  $d{=}4$ fermions  in $\brep{R}$, $\rep{R}$ of $\Lg$     &  calligraphic for anticommuting  \\
\hline
\smallstr  B_e \,\dd X^e       &   $\Lg$-valued  connection on $\IR^{3,1}$ &  occasionally embed in $A_{\Le_8}$\\
\hline
\smallstr  A_m\, \dd x^m       &   $\Lh$-valued connection for $\ccE$ on $\ccX$  &  occasionally embed in $A_{\Le_8}$\\
\hline
\smallstr \d \A         &  fluctuation of connection  for $\ccE$     &  occasionally  $\d A_{\Lh}$\\
\hline
\smallstr \d B         &  fluctuation of  connection  for  $\Lg$     &  occasionally use $\d A_{\Lg}$\\
\hline
\smallstr \ve         &  Majorana-Weyl $\so(9,1)$ spinor    & \\
\hline
\smallstr \z\otimes\l         &   $\so(3,1)\oplus\so(6)$ spinors   & $\l, \l'$ positive/negative chirality\\
\hline
\end{tabular}
\capt{5.0in}{notation}{A table of objects used. }
\end{center}
\end{table}

\begin{table}[H]
\def\bigstr{\vrule height22pt depth12pt width0pt}
\def\medstr{\vrule height19pt depth10pt width0pt}
\def\smallstr{\vrule height17pt depth9pt width0pt}
\begin{center}
\begin{tabular}{| >{\hskip2pt} l <{} >{$}l<{$} | >{$\hskip-3pt}c<{\hskip-3pt$} | 
>{$\hskip-3pt}c<{\hskip-3pt$} |}
\hline
\multispan2{\medstr\vrule\hfil Coordinates\hfil\vrule}&~\text{Holomorphic Indices}~&~\text{Real Indices}~\\ 
\hline\hline
%
\smallstr \cym                     & x^\m                & \m,\,\n,\ldots   & m,n,\ldots \\
\hline
\smallstr $\IR^{3,1}$ spacetime                 & X^e                 &   - & e,\,f,\ldots \\
\hline
\smallstr  basis for rep $\rep{r}$ of $\Lh$ ~~e.g. $\Lh = \su(3)$             &      [T_\Lh]^i{}_j  \in \rep{r}       &  i,j=1,\ldots, {r}      & - \\
\hline
\smallstr  basis for rep $\rep{R}$ of $\Lg$ ~~e.g. $\Lg = \Le_6$             &      [T_\Lg]^{M}{}_N  \in \rep{R}       &  M,N=1,\ldots,{R}      & - \\
\hline
\smallstr  parameters of heterotic structure   & y^\a  &\a,\,\b, \g,\ldots      & a,\,b,c\ldots \\
\hline
\smallstr  indices  for  $d=4$ spinors  (occasional) & \zeta_a, \zeb^\adot   & -      & a,b,\\
 \hline
\end{tabular}
\capt{5.0in}{notation2}{A table of coordinates and indices.}
\end{center}
\end{table}

\newpage
\section{Heterotic geometry}
\label{s2}
 The purpose of this section is to establish conventions and notation through a review of heterotic moduli geometry, most of which is explained in \cite{Candelas:2016usb}.  In terms of notation, there are occasional refinements and new results towards the end of the section. We largely work in the notation of \cite{Candelas:2016usb}, with a few exceptions, most important of which is that real parameters are denoted by $y^a$ and complex parameters by $y^\a, y^\bb$. The discussion both there, and in this section, refer to forms defined on the manifold $\ccX$. This is generalised in later sections in order to account for the charged matter fields. A table of notation given in tables \ref{notation}--\ref{notation2}. Basic results and a summary of conventions are found in the Appendices.  Hodge theory and forms are in Appendix \sref{app:Conventions}; spinors in Appendix \sref{app:Spinors}; and representation theory Appendix \sref{app:Representations}.

We consider a geometry $\IR^{3,1}\times \ccX$ with $\ccX$ smooth, compact, complex and vanishing first Chern class. While $\ccX$ is not \K in general, we take it to be cohomologically \K satisfying the $\del\delb$-lemma, meaning its cohomology groups are that of a \cym. 

The heterotic action is fixed by supersymmetry up to and including $\alpha'^2$--corrections. In string frame with an appropriate choice of connection for $\ccT_\ccX$, it takes the form \cite{Bergshoeff:1989de,Bergshoeff:1988nn}:
\begin{equation}
S = \frac{1}{2\kappa_{10}^2} \int\! \dd^{10\,}\! X \sqrt{g_{10}}\, e^{-2\Phi} \Big\{ \cR -
\half |H|^2  + 4(\del \Phi)^2 - \frac{\alpha'}{4}\big( \tr |F|^2 {-} \tr |R(\Theta^+)|^2 \big) \Big\} ~,
\label{eq:10daction}
\end{equation}
Our notation is such that  $\m,\n,\ldots$ are holomorphic indices along $\ccX$ with coordinates $x$; $m,n,\ldots$ are real indices along $\ccX$; while $e,f,\ldots$ are spacetime indices corresponding to spacetime coordinates $\ccX$.   The 10-dimensional Newton constant is denoted by $\kappa_{10}$, $g_{10}=-\det(g_{MN})$, $\Phi$ is the 10-dimensional dilaton, $\cR$ is the Ricci scalar evaluated using the Levi-Civita connection and $F$ is the Yang--Mills field strength with the trace taken in the adjoint of the gauge group. 

We define an inner product on $p$-forms by
$$
\langle S,\, T\rangle~=~ 
\frac{1}{p!} \, g^{M_1 N_1} \ldots g^{M_p N_p}\, S_{M_1\ldots M_p} \,T_{N_1 \ldots N_p}~.
$$
and take the $p$-form norm as
$$
|T|^2 ~=~\langle T,\, T\rangle~.
$$
Thus the curvature squared terms correspond to
$$
\tr |F|^2 = \half \tr F_{MN} F^{MN}~~~\text{and}~~~ \tr |R(\Th^+)|^2 ~=~
\half \tr R_{MNPQ}(\Theta^+) R^{MNPQ}(\Theta^+)~,
$$
where the Riemann curvature is evaluated using a twisted connection
$$
\Theta^\pm_M = \Theta_M \pm \half H_M~,
$$
with $\Theta_M$ is the Levi-Civita connection.
The definition of the $H$ field strength and its gauge transformations are given in \sref{sec:BHtransfs}.

We write the metric on $\ccX$ as
 $$
 \dd s^2 ~=~ 2g_{\m\nb} \dd x^\m \dd x^\nb~.
 $$
 The manifold $\ccX$ has a holomorphic $(3,0)$-form 
$$
\O = \frac{1}{3!} \O_{\m\n\r} \dd x^\m \dd x^\n \dd x^\r~,
$$
where $\O_{\m\n\r}$ depends holomorphically on parameters and coordinates of $\ccX$. $\O$ is a section of a line bundle over the moduli space, meaning there is a gauge symmetry in which $\O \to \mu \O$ where $\mu(y^\a) \in \IC^*$ is a holomorphic function of parameters. 

There is a compatibility relation
\beq
\frac{\ii}{ ||\O||^2} \O\, \Ob  ~=~ \frac{1}{3!}\o^3~, \qquad ||\O||^2 ~=~ \frac{1}{3!} \O_{\m\n\r} \Ob^{\m\n\r}~, 
\eeq
where $\ON$ is the norm of $\O$. For fixed $\ccX$, this is often normalised so that $||\O||^2 =8$.  However $\ON$ depends on moduli, and is gauge dependant and so it is not consistent in moduli problems to do this.

\subsection{Derivatives of $\O$ and $\D_\a$}

A variation  of complex structure given by a parameter $y^\a$ can be described in terms of the variation of the holomorphic three form by noting that $\del_\a\O\in H^{(3,0)}\oplus H^{(2,1)}$ and writing
\beq
\pd{\O}{y^\a}~=~-k_\a\,\O + \ch_\a~;~~~
\ch_\a~=~\frac12 \ch_{\a\,\k\l\bar\n}\,\dd x^\k \dd x^\l  \dd x^{\bar\n}~.\label{Kodaira}
\eeq
Here $\chi_\a$ are $\delb$-closed $(2,1)$-forms. Variations of complex structure $J:\ccT_\ccX \to \ccT_\ccX$ can also be phrased in terms of $(0,1)$-forms valued in $\ccT_\ccX$ 
$$
\del_\a J ~=~ 2\ii \,\D_{\a\,\nb}{}^\m \dd x^\nb \otimes \del_\m~, \qquad \del_\a \del_\b J ~=~  2\ii\,\D_{\a\b} ~=~  2\ii\,\D_{\a\b}{}_\nb{}^\m \dd x^\nb \otimes \del_\m~.\notag
$$
We have denoted $\del_\b \D_\a = \D_{\a\b}$ which makes manifest the symmetry property $\del_\b \D_\a = \del_\a \D_\b$. Occasionally we will denote parameter derivatives by $\del_\a \o \cong \o_{,\a}$.

The $\D_\a$ and $\chi_\a$  are related
\beq
\chi_\a ~=~ \frac{1}{2}\O_{\r\s}^{~~~\nb}  \D_{\a\,\mb\nb}  \dd x^\r \dd x^\s \dd x^\mb~, \qquad \D_\a^{~\m} ~=~ \frac{1}{2\norm{\O}^2} \Ob^{\m\t\r} \chi_{\x\,\t\r\,\sb} \, \dd x^\sb~.
\label{eq:DeltaDef}\eeq
The symmetric component of $\D_\a{}^\m$ appears in variations of the metric $\d g_{\bar\m\bar\n}=\D_{\a\,(\bar\m\bar\n)}\, \d y^\a$. 

It is best to describe variations of complex structure through projectors $P$ and $Q$ onto holomorphic and antiholomorphic components respectively
$$
P_m{}^n ~=~ \half (\d_m{}^n - \ii J_m{}^n)~, \qquad Q_m{}^n ~=~ \half (\d_m{}^n + \ii J_m{}^n)~.
$$
The projectors capture the implicit dependence on complex structure. For example, the operator
$
\delb~=~\dd x^m Q_m{}^n\, \partial_n
$
undergoes a variation purely as a consequence of the implicit dependence on the complex structure: 
\beq
[\del_\a,\delb] ~=~[ \del_\a, (Q_m{}^n \dd x^m \otimes \del_n)] ~=~ - \D_\a{}^\m \del_\m
\label{derivdelbar}\eeq
%


\subsection{The vector bundle $\ccE$}
Let $\ccE$ denote a vector bundle over $\ccX$, with structure group $\cH$, and $A$ the connection on the associated principal bundle. That is, $A$ is a gauge field valued in the adjoint representation $\ad_\Lh$ of the Lie algebra $\Lh$ of $\cH$. 

Under a gauge transformation, $A$ has the transformation rule
\beq
^{\Phi}A = \Phi (A - Y)\Phi^{-1}~, \quad Y=\Phi^{-1}\dd \Phi~,\label{eq:GaugeTransf1}
\eeq
where $\Ph$ is a function on $\ccX$ that takes values in $\eu{G}$. We take $\Ph$ to be unitary and then $ \dd\Ph\,\Ph^{-1}$ and $A$ are  antihermitean. The field strength is
$$
F~=~\dd A + A^2~,
$$
and this  transforms in the adjoint of the gauge group: $F\to \Ph F\Ph^{-1}$.

Let $\A$ be the $(0,1)$ part of $A$ then, since $A$ is antihermitean,
$$
A = \A - \A^{\dag}~.
$$
On decomposing the field strength into type, we find $F^{0,2} = \delb\A + \A^2$.  
The bundle $\ccE$ is holomorphic if and only if there exists a connection such that $F^{(0,2)}=0$. The Hermitean Yang-Mills equation is
$$
\o^2 F ~=~0~.
$$

\subsection{The $B$ and $H$ fields\label{sec:BHtransfs}}
There is a gauge-invariant three-form
\beq
H~=~\dd B - \frac{\ap}{4}\Big(\CS[A] - \CS[\Th]\Big)~,
\label{Hdef}\eeq
where $\CS$ denotes the Chern-Simons three-form
$$
\CS[A]~=~\tr\!\left(A\dd A +\smallfrac23\, A^3\right)~=~\tr\!\left(AF - \smallfrac13\, A^3\right)~,
$$
and $\Th$ is the connection on $\ccT_\ccX$ for Lorentz symmetries. The three-form $\dd B$ is defined so that  $H$ to be gauge invariant, and so $\dd B$ itself has gauge transformations. 
Under a gauge transformation 
$$
\CS[A]~\to~\CS[A] - \dd\tr\!(AY) + \frac13 \tr\left(Y^3\right)~,
$$
together with the analogous rule for $\CS[\Th]$. 
The integral of $\tr(Y^3)$ over a three-cycle is a winding number, so vanishes if the gauge transformation is continuously connected to the identity. The integral vanishes for every three-cycle and so
$\tr(Y^3)$ is exact
$
\frac13\,\tr(Y^3) ~=~\dd U~,
$
for some globally defined two form $U$. There are corresponding transformations for the connection $\Th$ in which $Y$ is replaced by $Z$ and $U$ by $W$.

Anomaly cancellation condition means that the $B$ field is assigned a transformation
\beq\label{eq:BTransf}
B~\to~B - \frac{\ap}{4}\left\{\tr(AY) - U - \tr(\Th Z) + W\right\}~.
\eeq
With this transformation law, $B$ is a 2-gerbe and $H$ is invariant.

An important constraint arising from supersymmetry is that $H$ is related to the hermitian form $\o$ and complex structure $J$ of $\ccX$:
\beq
H ~=~  \dd^c\o~, \qquad \dd^c \o  ~=~ \half J_{m_1}{}^{n_1}J_{m_2}{}^{n_2} J_{m_3}{}^{n_3} (\del_{n_1}\o_{n_2n_3})\, \dd x^{m_1} \dd x^{m_2} \dd x^{m_3}~,\label{eq:Hsusy}
\eeq
which for an integrable complex structure reduces to 
\beq
\dd^c \o ~=~ J^m \del_m \o - (\dd J^m) \o_m~.\label{eq:dco}
\eeq
We denote the real parameters of the compactification by $y^a$ and complex parameters by $y^\a,y^\bb$.  If the parameters are fixed to $y=y_0$ the second term in \eqref{eq:dco} vanishes and the relation simplifies to 
\beq
\dd^c \o|_{y=y_0} ~=~ \ii (\del -  \delb) \o~.\label{eq:dco_old}
\eeq
However, when complex structure is varied $\del_\a (\dd J) = 2\ii \del \D_\a$ the second term in \eqref{eq:dco} is non-zero and is important as it contributes to the equations satisfied by the moduli.

\subsection{Derivatives of $A$}
The heterotic structure  $(\ccX,H,\ccE)$ depends on parameters.  This means the gauge connection $A$ and its gauge transformations $\Phi$  depend on parameters. As constructed in \cite{Candelas:2016usb}, the gauge covariant way of describe a deformation of $A$ is given by introducing a covariant derivative
\beq
D_a A~=~\partial_a A - \dd_A A_a~,
\label{CovA}\eeq
where $A^\sharp = A_a \dd y^a$ is a connection on the moduli space with a transformation law 
\beq\label{LambdaRule}
{}^\Phi A_a ~=~\Ph (A_a - Y_a)\Ph^{-1} ~,\qquad  Y_a ~=~ \Ph^{-1}\,\partial_a \Ph ~.
\eeq
With this transformation property $D_a A$  transforms homogeneously under \eqref{eq:GaugeTransf1}:
\beqnn
D_a A~\to~ \Ph \,D_a A\,\Ph^{-1}~.
\eeqnn
The moduli space $\cM$ is complex, and we introduce a complex structure $y^a = (y^\a, y^\bb)$. 
When parameters vary complex structure the holomorphic type of forms change, and the covariant derivatives $D_\a \A$ is no longer gauge covariant. This is remedied by defining a generalisation, termed the holotypical derivative $\ccD_\a \A$:
\beq\begin{split}
\ccD_\a\cA~&=~(D_\a A)^{(0,1)}~=~\partial_\a\cA - \D_\a{}^\m \cA^\dag_\m - \delbA A^{\sharp}_\a~, \\[5pt]
\ccD_{\bb}\cA~&=~(D_{\bb}A)^{(0,1)}~=~
\partial_{\bb}\cA - \D_{\bb}{}^{\bar\m} \cA_{\bar\m} - \delbA A^\sharp_{\bb}~=~0~,
\end{split}\label{CpxCovA}\eeq
where the vanishing of $\ccD_{\bb}\cA$ follows from \eqref{derivdelbar}. It follows from the definition 
that under a gauge transformation the holotypical derivative transforms in the desired form
\beq
\ccD_\a\cA ~\to~ \Ph\, \ccD_\a\cA\, \Ph^{-1}~.
\notag\eeq
Without the extra term $- \D_\a{}^\m \cA^\dag_\m$ in the holotypical derivative, this  property does not hold as $\delb$ fails to commute with $\del_\a$.

The holotypical derivative can be extended to act on $(p,q)$-forms. Define
$$
W_m^{r,s} ~=~ \frac{1}{r!s!}\, W_{m \m_1\cdots\m_r \nb_1\cdots\nb_s} 
\dd x^{\m_1\cdots\m_r\nb_1\cdots\nb_s}~,
$$
and understand $W_m^{r,s} = 0$ if $r$ or $s$ are negative or $r{+}s>n{-}1$.  The holotypical derivatives are then given by
\beq\begin{split}
\ccD_\a W^{p,q} ~&=~(D_\a W)^{p,q}~=~D_\a W^{p,q} - \D_\a^{~\m} W_\m^{p-1,q} + \D_\a^{~\m} W_\m^{p,q-1}~,\\[8pt]
\ccD_{\bb} W^{p,q} ~&=~(D_{\bb} W)^{p,q}~=~D_{\bb} W^{p,q} + 
\D_{\bb}{}^{\bar\n} W_{\bar\n}^{p-1,q} - \D_{\bb}{}^{\bar\n} W_{\bar\n}^{p,q-1}~.
\end{split}\label{eq:CovariantDerivW}\eeq
The holotypical derivative has the nice feature that it preserves holomorphic type:
\beq
\ccD_\bb \ccD_\a W^{p,q}~=~(D_\bb \ccD_\a W)^{p,q}~=~(D_\bb D_\a W)^{p,q}~.
\notag\eeq
We use $D_\a$ to denote the covariant derivative to account for any gauge dependence of the real form $W$. For example, the covariant derivative of the field strength is related to that of $A$:
\beqnn
D_M F~=~\partial_M F + [ A^\sharp_M, F] ~=~ \dd_A (D_M A),
\eeqnn
However, the holotypical derivative of, say $F^{(0,2)}$, gives 
\beq
\ccD_\a F^{(0,2)}~=~ \delb_\A \,\ccD_\a \A ~=~ \D_\a{}^\m F_\m~,
\notag\eeq
and this is known as the Atiyah constraint.

\subsection{Derivatives of $H$}
It is of use to compute derivatives of $H$ with respect to parameters. First  define a gauge covariant derivative of $B$ via
\beq
D_a B = \del_a B -\frac{\ap}{4} \tr ( A_a\,\dd A) ~,
\label{eq:CovDerivB}
\eeq
With this choice, we have a gauge transformation law for~$D_a B$ that is parallel to the gauge transformation \eqref{eq:BTransf} for $B$: 
\beq
 {}^\Phi D_a B ~=~ D_a B + \frac{\ap}{4}\Big(\! \tr( Y D_a A ) + U_a \Big) ~.
\label{eq:covBTransf}\eeq
The second and third derivatives are defined to transform in a natural way inherited from that of $D_a B$ 
\beq
\begin{split}
 D_a D_b B ~&=~  \del_a D_b B - \frac{\ap}{4} \tr ( D_b A  \dd A_a)~,\cr
 D_cD_a D_b B ~&=~  \del_c D_a  D_b B - \frac{\ap}{4} \tr ( D_b D_a A\, \dd  A^\sharp_c).
\label{eq:SecondDerivB}
\end{split}
\eeq

A gauge invariant quantity $\ccB_a$ is the formed from $D_a B$
\beq
\ccB_a ~=~ D_a B + \frac{\ap}{4} \tr (A D_a A) -\dd b_a~,
\label{eq:curlyBdef}\eeq
with $\dd b_a$ an exact form. The exact form comes from the fact the physical quantity is $\dd B$, and so in writing $\ccB_a$ there is a corresponding  ambiguity. 
It is a simple exercise to note that $\del_a H$ is given by the expression
\beq 
\del_a H ~=~  \dd \ccB_a - \frac{\ap}{2} \tr ( D_a A\, F ) ~.
\label{eq:HDeriv}\eeq
In terms of holomorphic parameters, we introduce the holotypical derivative and find
\beq
\ccD_\a H^{p,q}~=~\del\ccB_\a^{p-1,q} + \delb\ccB_\a^{p,q-1} - 
\frac{\ap}{2} \tr\!\big(\ccD_\a\cA\, F^{p,q-1}\big)~.
\label{eq:varHByType}\eeq

The second derivative is given by 
\beq
\begin{split}
 \del_b \del_a H ~&= - \frac{\ap}{2} \tr \Big( (\dd_A D_b A) D_a A  + F D_b D_a A \Big)  + \dd\ccB_{ba}~,
\end{split}\label{eq:Hsecondderiv}
\eeq
where $\ccB_{ba} = \del_b \ccB_a$ and in terms of the B-field
$$
\ccB_{ba} ~=~ \left( D_{b} D_{a} B - \frac{\ap}{4} \tr \big((D_b D_a A ) A \big) \right)~.
 $$
Despite appearances the right hand side is symmetric in $a,b$ after one uses that
\beq
D_{[b} D_{a]} B  = - \frac{\ap}{4} \tr \dd A \IF_{ba} \frac{\ap}{4} \dd \tr [ A_a,  A^\sharp_b] A~.
\eeq
and so 
\beq
\begin{split}
  \del_b \del_a H ~&= - \frac{\ap}{4} \tr \Big(( \dd_A D_b A) D_a A + (\dd_A D_a A) D_b A + F\{  D_{a} ,D_{b}\} A \Big)   \\[5pt]
 &\quad + \dd \left( D_{(b} D_{a)} B - \frac{\ap}{4} \tr \big((D_{(b} D_{a)} A ) A \big) \right)~.
\end{split}
\eeq

The third derivative of $H$ is given by 
\beq
\begin{split}
 \del_c \del_b \del_a H~&= - \frac{\ap}{2} \tr \Big(\{ D_c A, D_b A\} D_a A+ (\dd_A D_c D_b A) D_a A + (\dd_A D_b A) D_c D_a A  \\[5pt]
 &\qquad\qquad\qquad + (\dd_A D_c A) D_b D_a A  \Big) +  \dd \ccB_{cba}~.
\end{split}\label{eq:thirdderiv}
\eeq
where $\ccB_{cba} = \del_c \del_b \ccB_a$ and in terms of the B-field:
$$
\dd \ccB_{cba} ~=~  \left( D_c D_{b} D_{a} B - \frac{\ap}{4} \tr \Big((D_c D_b D_a A ) \,A + (D_b D_a A)\, (D_c A)  \Big) \right)~.
$$
For similar reasons to the second derivative above, this is actually symmetric in $a,b,c$, but not made manifest in this expression for compactness. 

\subsection{Derivatives of $\dd^c \o$}

The derivative of $\dd^c \o$ in \eqref{eq:dco} with respect to parameters is
\beq
\begin{split}
 \del_\a\, (\dd^c \o)  ~=&~ 2\ii \D_\a{}^m \del_m \o + J^m \del_m \del_\a \o_m - 2\ii \,(\dd \D_\a{}^m) \o_m - (\dd J^m) \del_\a \o_m~.\\[2pt]
\end{split}
\label{eq:dcomDeriv}
\eeq
We can evaluate \eqref{eq:dcomDeriv} for a given complex structure, denoted $|_{y_0}$ in the corresponding complex coordinates of $\ccX$:
\beq
 \del_\a\, (\dd^c \o)|_{y_0} ~=~ 2\ii \D_\a{}^\m (\del_\m \o - \del \o_\m) + \ii (\del-\delb) \del_\a \o  - 2\ii \del( \ccD_\a \o^{0,2})\notag
\eeq
For a given complex structure, we project onto holomorphic type then we need to use holotypical derivatives. Two cases we will need and then projecting onto $(0,3)$ and $(1,2)$ components 
\beq\label{eq:dcderiv}
\begin{split}
 (\del_\a \dd^c \o)|_{y_0}^{0,3} ~&= -\ii \delb \ccD_\a \o^{0,2}, \\[3pt]
 (\del_\a \dd^c \o)|_{y_0}^{1,2} ~&=~ 2\ii \D_\a{}^\m (\del_\m \o-\del\o_\m) - \ii \del \,\ccD_\a \o^{0,2} -\ii \delb\, \ccD_\a \o^{1,1}~.\\[2pt]
\end{split}
\eeq

The second derivative of $\dd^c\o$, given by differentiating \eqref{eq:dcomDeriv} and then evaluated on a fixed complex structure, is
\beq
\begin{split}
( \del_\a\del_\b \dd^c\o)|_{y_0} &=~  2\ii ( \D_{\a\b}{}^\m)(\del_\m \o)  + 2\ii \D_\a{}^\m \del_\m (\del_\b \o) + 2\ii \D_\b{}^\m \del_\m (\del_\a \o)   - 2\ii \,(\del \D_{\a\b}{}^\m) \o_\m  \\[3pt]
 &\quad+ \ii(\del-\delb) \o_{,\a\b} -2\ii (\del \D_\a{}^\m) \o_{\m,\b} - 2\ii (\del \D_\b{}^\m) \o_{\m,\a}~.
\end{split}\notag
\eeq
We will have need for the $(0,3)$-component:
\beq
\begin{split}
\Big( \del_\a\del_\b \dd^c\o\Big)|_{y_0}^{0,3} ~&=~  2\ii( \D_\a{}^\m  \D_\b{}^\n + \D_\b{}^\m\D_\a{}^\n)\del_\m \o_\n^{0,1}   -\ii \delb  ( \o_{,\a\b})^{0,2}~.\label{eq:dc03}
\end{split}
\eeq

\subsection{Supersymmetry relations}
One can apply these results to compute how the supersymmetry condition $\dd^c \o = H$ relates the variations of fields. The parameter space coordinates are corrected at order $\ap$. Differentiating with respect to these corrected coordinates, the equation \eqref{eq:Hsusy} gives rise to relations between first order deformations of fields:
\beq
\begin{split}
\ccB_\a^{2,0}  ~=~ \del \b^{1,0},~~&~~~~~ \ccD_\a\o^{2,0} ~=~ 0~,\\[3pt]
 \ccB_\a^{0,2} + \ii \ccD_\a \o^{0,2} ~&=~ \delb \k_\a^{0,1}~, \\[3pt] 
\ccB_{\a}^{1,1} - \ii \ccD_\a \o^{1,1} &=~ 0~,\\[3pt]
  2\ii \D_\a{}^\m (\del_\m \o - \del \o_\m) + \frac{\ap}{2} \tr( \ccD_\a \A F) ~&=~ \delb (\ccB_\a^{1,1} + \ii \ccD_\a\o^{1,1} -  \del \k_\a^{0,1})~. 
\end{split}\label{eq:cBrelations1}
\eeq
where $\g_\a^{1,1}$ is $\dd$-closed $(1,1)$-form, and $k_\a^{0,1}$ and $l_\a^{0,1}$ are some $(0,1)$-forms. As discussed in \cite{Candelas:2016usb} in $\ap$-perturbation theory, $\ccB_\a^{0,2} = \ccD_\a \o^{0,2} = \cO(\ap)$ when appropriately gauge fixed.  

The heterotic structure are holomorphic functions of parameters. This can be compactly stated as
\beq
\D_\ab{\,}^\n \del_\n ~=~ 0~, \quad \ccD_\ab \A ~=~ 0~, \quad \ccB_\ab^{1,1} + \ii \ccD_\ab \o^{1,1} ~=~ 0~.
\eeq

\newpage
\section{The matter field metric}
\label{s3}
 In this section we dimensionally reduce the Yang-Mills term in \eqref{eq:10daction} to obtain the metric for the matter fields. 
 Our task divides into two steps. First, determine how $A_{\Le_8}$ decomposes under  $\Le_8\oplus\Le_8\supset \Lg \oplus \Lh$, using this to form a KK ansatz. For simplicity we will suppress writing the second $\Le_8$ sector. Second, use this  to dimensionally reduce the $d=10$ action thereby getting an effective field theory metric and Yukawa couplings for the matter fields and construct the \K potential.  

\subsection{Decomposing $A$ under $ \Lg\oplus \Lh\subset \Le_8$}\label{sect:ADecomp}

The branching rule for $ \Lg \oplus \Lh \subset \Le_8$ is
\beq\label{eq:branching}
{\rm ad}_{\Le_8} ~=~ (\rep{1}, {\rm ad}_{\Lh}) \oplus ({\rm ad}_{\Lg}, \rep{1}) \oplus_i (\brep{R_i}, \rep{r_i}) \oplus(\rep{R_i}, \brep{r_i})~.
\eeq
where $\ad$ denotes the relevant adjoint representation. The matter fields transform in representations of $\Lg \oplus \Lh$. We denote these representations by $\rep{r_i}$ for $\Lh$ and  $\rep{R_i}$ of $\Lg$ respectively. Denote dimensions in the obvious way $\dim \rep{r_i} = r_i$ and $\dim \rep{R_i} = R_i$.  We have allowed for a sum over all the relevant representations $\rep{r_i}$ and $\rep{R_i}$, including for example pseudo-real representations. For simplicity we will often suppress the sum and write a single matter field representations of $\Lg\oplus \Lh$ and its conjugate. The generalisation is obvious.

The matrix presentation of the adjoint of $\Le_8$ is complicated. For the moment let us suppose we can write the generator in simplified form as 
\beq
T_{\Le_8} = 
\begin{pmatrix}
 T_{\Lh} & 0\\
 0 & T_{\Lg}
\end{pmatrix} ~.
\eeq
We do this as a toy model to illustrate the key points of the calculation, and at the end of the day our results will not depend on this presentation. 

The background gauge field is 
$$
A_{\Le_8} ~=~ A_\Lh \oplus B_\Lg~,
$$
 where $A_\Lh = A_m(x) \dd x^m$ is the $\Lh$ gauge field and we take it to have legs purely along on the CYM; $B_\Lg = B_e(X) \dd X^e$ is the $\IR^{3,1}$ spacetime gauge field valued in  $\Lg$. When there is no ambiguity we will drop the  $\Lg, \Lh$ subscripts. We indicate this combined Lorentz and gauge structure schematically in matrix notation
$$
A_{\Le_8} = 
\begin{pmatrix}
 A_{\Lh} & 0 \\
 0 & B_{\Lg} \\
\end{pmatrix}  = A_m \dd x^m + B_e \dd X^e~.
$$
Note that having off-diagonal terms turned on in the background would amount to Higgsing the gauge group, which we do not want for the discussion in this paper.

A fluctuation of $A_{\Le_8}$ is  of the form
\beq
\begin{split}
 \d A_{\Le_8} ~&=~ \d A_{(\ad_\Lh,\rep{1})} \oplus_i  \d A_{(\rep{r_i},\brep{R_j})}\oplus_j  \d A_{(\brep{r_j},\rep{R_j})} \oplus \d A_{(\rep{1}, \ad_\Lg)} \\
&=~ \begin{pmatrix}
 \d A_{\Lh} & \Phi \\
 \Psi & \d B_{\Lg} \\
\end{pmatrix},
\qquad
\Phi_{r\times R} = 
\begin{pmatrix}
 \Phi^1 &\cdots &
 \Phi^R
\end{pmatrix},
\qquad
\Psi_{R\times r} = 
\begin{pmatrix}
 \Psi^1 \\ 
 \vdots\\
 \Psi^R
\end{pmatrix},
\end{split}
\label{eq:Afluct}
\eeq
where again we have used matrix notation to indicate the structure. This gives us an intuition for understanding the transformation properties under a  $\Lh\oplus\Lg$ rotation:
\beq
\begin{split}
\d A_{\Le_8} ~&\to~ \d A_{\Le_8} + [T_{\Le_8} , \,\d A_{\Le_8}] \\[5pt]
&=~ 
\d A_{\Le_8}  + \begin{pmatrix}
[  T_{\Lh},\, \d A_\Lh] & T_{\Lh} \,\Phi - \Phi\, T_{\Lg} \\
 T_{\Lg}\, \Psi - \Psi\, T_{\Lh} &  [ T_{\Lg}, \,\d B_\Lg ] ~.
\end{pmatrix}
\end{split}
\eeq
We see that fields transform under $\Lg \oplus \Lh$ as:
\begin{enumerate}
 \item $\d A_\Lh$ transforms as $(\rep{1},\ad_{\Lh})$, $\d B_\Lg$ transforms as $(\ad_\Lg, \rep{1})$; 
 \item $\Phi$ is in the $(\brep{R},\rep{r})$ and is a $r\times R$ matrix. Column vectors are in the fundamental; row vectors the antifundamental. For example,  $\Phi^{1},\cdots,\Phi^R$  are each column vectors transforming in the $\rep{r}$ of $\Lh$.
 \item   $\Psi$ is in the $(\rep{R},\brep{r})$ and is a $R\times r$ matrix with $\Psi^1,\ldots,\Psi^R$ row vectors and so in the $\brep{r}$ of $\Lh$. 
 \item To preserve the structure $\Lg \oplus \Lh$, $\d A_\Lh$ has legs only on  the CYM while $\d B_\Lg$ has legs only in $\IR^{3,1}$.  
 
\end{enumerate}

The reality condition is $\d A_{\Le_8}^\dag = - \d A_{\Le_8}$ which implies
$$
\d A_\Lh^\dag = - \d A_\Lh~, \qquad \Phi^\dag = - \Psi~, \qquad \d B_\Lg^\dag = - \d B_\Lg~.
$$
Decomposing this condition according to the holomorphic type of $\ccX$ we have:
$$
\d A_\Lh = \d \A - \d \A^\dag~, \qquad \Phi = \phi- \psi^\dag~, \quad  \Psi =\psi - \phi^\dag~, 
$$
where 
$$
\phi= \Phi_{0,1}~, \quad \psi = \Psi_{0,1}~,
$$
while the $(1,0)$-components are fixed by reality: $\Phi_{1,0} = -\psi^\dag$, $\Psi_{1,0} = -\phi^\dag$.\footnote{It may be useful to define $\bar \Psi_\m^a$ given by $ \bar \Psi_\m^a =( \Psi_\mb^a)^* $ so that $a$ is a real index and $\Phi_{1,0}^a = - \psi^{*\,a} = - \bar\Psi^a_{1,0} $.}

The $\Le_8$ field strength is 
$$
F_{\Le_8} = \dd A_{\Le_8} + A_{\Le_8}^2~.
$$
Its background value decomposes according to its orientation of legs:
\beq
\begin{split}
 F_{\Le_8 ef} \,\dd X^e \,\dd y^f ~&=~\dd B_\Lg + B_\Lg^2~,\cr
  F_{\Le_8 mn} \,\dd x^m \,\dd x^n ~&=~ \dd A_\Lh + A_\Lh^2~,\cr
   F_{\Le_8 me} \,\dd x^m \,\dd X^e ~&=~  \left(\del_m B_{\Lg\,e} - \del_e A_{\Lh\,m} +   A_{\Lh\,m} B_{\Lg\,e} - B_{\Lg\,e} A_{\Lh\,m}\right)\,\dd x^m \,\dd X^e = 0~.\cr
\end{split}
\eeq
Under  $A_{\Le_8} \to A_{\Le_8} + \d A_{\Le_8}$,
\beq
\begin{split}
 \d F_{\Le_8} ~&=~ \dd_{A_{\Le_8}} \d A_{\Le_8} \\[2pt]
&=~ \begin{pmatrix}
 \dd_A( \d A_\Lh) & \dd \Phi + A_\Lh \Phi  + \Phi B_\Lg\\
 \dd \Psi +  \Psi A_\Lh + B_\Lg \Psi & \dd_{B_\Lg} (\d B_\Lg   )
 \end{pmatrix}~. \\[5pt]
\end{split}
\label{eq:Ffluct}
\eeq

Consider $\d F_{\Le_8}$ oriented along $\ccX$. At this point we drop the $\Lh, \Lg$ subscripts on $A, B$.  Then, the equations of motion require $F_{\Le_8}$ be $(1,1)$ implying any $(0,2)$-component must satisfy,
$$
\d F^{0,2}_{\Le_8}\big|_{\ccX} = 
\begin{pmatrix}
 \delb_\A( \d \A)  & \delb \phi+ \A\phi\\
 \delb\psi + \psi\A & 0
\end{pmatrix} = \begin{pmatrix}
 \d y^\a  \D_\a^{~\m} F_{\Lh\,\m}^{\; 0,1}  & 0\\
0 & 0  \\
\end{pmatrix}~.
$$

The off-diagonal terms tell us the fields $\phi,\psi$ are holomorphic sections of $\ccE$. We will occasionally introduce index to make manifest the fact $\phi$  is in the $(\brep{R}, \rep{r})$ of $\Lg \oplus \Lh$ by writing $\phi^{i\,\Mb}$ where $i,\jb=1, \cdots,  {r}$ for representations $\rep{r}$ of $\Lh$; $M,\Nb=1,\cdots, R$ for representations $\rep{R}$ of $\Lg$. Then, for example, 
\beq
\begin{split}
\delb_\A \phi^\Mb  &=  \delb \phi^\Mb + (\A\phi)^\Mb = 0 \quad\longrightarrow\quad \phi^\Mb\in H^1(\ccX,\ccE_{\rep{r}})~, \cr
 \delb_\A\psi^N &=  \delb\psi^N + (\psi \A)^N= 0 \quad\longrightarrow\quad \psi^N\in H^1(\ccX,\ccE_{\brep{r}})~. 
\end{split}\label{eq:MatterCohomology}
\eeq
We have introduced the $\delb$-cohomology group $H^1(\ccX,\ccE_{\rep{r}})$, with forms valued in the  $\Lh$-subbundle of $\ccE$ whose fibres are the representation $\rep{r}$ of $\Lh$. The $\rep{r}$ index $i,\jb$ is implicitly summed.

We could also study the equations of motion for $\d A_{\Le_8}$ 
$
\dd^\dag_A (\dd_A \d A_{\Le_8}) = 0~.
$
Choosing the gauge $\dd_A^\dag \d A_{\Le_8} = 0$ we see
$
\Box_A \d A_{\Le_8} = 0~.
$
For example, on the $\phi$ matter field this gives
$$
(\delb_\A \delb_\A^\dag + \delb_\A^\dag \delb_\A ) \phi = 0~.
$$
This has solution if $\phi$ is harmonic element of $H^1(\ccX,\ccE_{\rep{r}})$. This is slightly stronger than the cohomology relation \eqref{eq:MatterCohomology}.

Expand the fields $\phi$ and $\psi$ in a harmonic basis for $H^1(\ccX,\ccE_{\rep{r}})$ and $H^1(\ccX,\ccE_{\brep{r}})$ respectively: 
\beq
\begin{split}
 \phi ~&=~ \sum_{\x} \phi_{\x} \,  C^{\x}\,\in (\rep{r},\brep{R})~, \qquad   \psi ~=~ \sum_{\t}  \psi_{\t} \, D^{\t} \, \in(\brep{r},\rep{R})~, \cr
\end{split}
\eeq
where $\phi_{\x}\in H^1(\ccX,\ccE_{\rep{r}})$ and $\psi_{\t} \in H^1(\ccX,\ccE_{\rep{r}})$ are harmonic forms  
\beq
\phi_{\x} = \phi_{\x\,\mb}\,\dd x^\mb \in \rep{r}~, \qquad \psi_{\t} = \psi_{\t\,\mb}\, \dd x^\mb \in \brep{r}~.
\eeq
while $C^\x$ and $D^\t$ are valued in $\brep{R}$ and $\rep{R}$ respectively. 

For example, consider the standard embedding. Then,  $\ccE_{\rep{3}} = \ccT^{1,0}_\ccX$ and $\phi_\x \in H^1(\ccX,\ccT^{1,0}_\ccX)$; $\ccE_{\brep{3}} = \ccT^{0,1}_\ccX$ with the $\psi_{\t}\in H^1(\ccX,\ccT^{0,1}_\ccX)$.    $C^\x$ and $D^\t$ are in the $\brep{R} = \brep{27}$ and $\rep{R} = \rep{27}$. 


We need to satisfy the reality condition $\Phi^\dag = -\Psi$, which forces $\phi^\dag =- \psi$ and so in terms of the $(\phi_{\x},\psi_{\t})$ basis:
\beq\label{eq:MatterKK}
\Phi  ~=~ \sum_{\x} \phi_{\x} \,C^{\x} - \sum_{\tb}  \psi_{\tb}\,D^{\,\tb} ~,\qquad 
\Psi  ~=~ \sum_{\t}  \psi_{\t} \,D^{\t} - \sum_{\xb}  \phi_{\xb} \,C^{\xb}\,.
\eeq
We denote conjugation through the barring of the indices. For example, $\phi_\xb = (\phi_\x)^\dag$ is a $(1,0)$-form valued in $\brep{r}$ of $\Lh$ and $C^\xb = (C^\x)^\dag$ is in the $\rep{R}$ of $\Lg$.

\subsection{The matter field metric from reducing Yang-Mills, $\cL_F$}
The spirit of KK reduction is to promote the  coefficients to spacetime fields: $Y^\a(X)$,  $C^{\x}(X),~D^{\t}(X)$, and integrate over the six-dimensional manifold to get an effective four-dimensional theory.   
With the conventions of \cite{Candelas:2016usb}, the $d=10$ $\Le_8$ Yang-Mills field contribution to the $d{=}4$ effective field theory is: 
\beq\label{eq:YM}
\cL_F ~=~ -\frac{\ap}{4V}\int_\ccX \dd^6 x \sqrt{g}\, \tr |\d F_{\Le_8}|^2 ~, \quad |F|^2 = \frac{1}{2} F_{MN} F^{MN}~.
\eeq
We dimensionally reduce, doing a background field expansion. A small fluctuation of the field strength is given by \eqref{eq:Ffluct}, and so 
\beq
\begin{split}
 \tr |\d F_{\Le_8}|^2 ~&=~  \tr \left(\dd_{A}( \d A )\,\star \,\dd_{A}( \d A) \right) + \tr \left(\dd_{A+B} \Phi\, \star\,\dd_{A+B} \Psi \right) \\[4pt]
&\quad+  \tr \left(\dd_{A+B} \Psi \,\star\, \dd_{A+B} \Phi \right) + \tr \left(\dd_{B} \d B \, \star\,\dd_{B} \d B \right)~,\label{eq:Ffluct2}
\end{split}
\eeq
The first term involves just the bundle moduli, contributing to the moduli metric considered in \cite{Candelas:2016usb}; the middle two terms involve the matter fields and the last term gives rise to the kinetic term for the $d{=}4$ spacetime gauge field. The terms involving the matter fields are:
\beq
\begin{split}
\dd_{A+B} \Phi ~&=~ (\del_e \Phi + \Phi B_e) \,\dd X^e + (\del_M \Phi_N + A_M \Phi_N)\, \dd x^M \dd x^N \\[5pt]
&=~ \widehat\cD_e \Phi \,\dd X^e+ \dd_A \Phi~,\\[5pt]
 \dd_{A+B} \Psi 
 &=~ \widehat\cD_e \Psi\, \dd X^e + \dd_A \Psi~,\\[5pt]
\end{split}
\eeq
where $\widehat\cD_e$ is the spacetime $\Lg$-covariant derivative and $\dd_A$ the $\Lh$-covariant derivative. 
Hence, using 
$
\tr |\d F|^2 ~=~ \half \tr \d F_{MN} \d F^{MN} ~=~ 2 \tr \d F_{e\m} \d F^{e\m} ~,
$
where $\tr (\d F_{e\m} \d F^{e\m} ) = \tr (\d F_{e\nb} \d F^{e\nb})$, and ignoring the moduli fields  for the moment, we find the  kinetic terms for the matter fields come from middle two terms in \eqref{eq:Ffluct2} and are
\beq
\begin{split}
\tr |\d F_{\Le_8}|^2 ~=~ - 2 \tr \left( \widehat\cD_e \Phi_\mb\, \widehat\cD^e \Phi^{\dagger\,\mb}\right)  - 2\tr \left( \widehat\cD_e \Psi_\mb   \, \widehat\cD^e \Psi^{\dagger\,\mb}\right)~.
\end{split}\label{eq:YMRed1}
\eeq

We have used the reality condition $\Phi^\dag = -\Psi$. 
The matter fields have a KK anstaz, given by \eqref{eq:MatterKK}, which when substituted into each of the above terms gives
\beq
\begin{split}
\vol \tr\left( \widehat\cD_e\Phi_\mb\, \widehat\cD^e \Phi^{\dagger\,\mb}\right) ~&=~\left(  \widehat\cD_eC^{\x\,\Nb}(X)\phi_{\x\,\mb}^{\,i}(x)\right) \left(\widehat\cD^eC^{\eb \,M}(X) \phi_{\eb}^{\,\mb \jb}(x)\right)  \d_{i\jb}\d_{M\Nb}\,(\star\, 1) \\[3pt]
&=~ \left(\widehat\cD_e C^{\eb\,M}(X)\right) \;\left(\widehat\cD^e C^{\x \,\Nb}(X) \right)\,\left(\phi_{\x}^{\,i} (x)\star \phi_{\eb}^{\jb}(x) \right)\,\d_{i\jb}  \d_{M\Nb}~,\\[5pt]
\vol \tr\left( \widehat\cD_e \Psi^{}_\mb   \, \widehat\cD^e \Psi^{\dagger\,\mb}\right) &=~\left(  \widehat\cD_e D^{\,\s\,M}(X)\psi_{\s\,\mb}^{\,\jb}(x) \right) \left(\widehat\cD^e D^{\tb\,N}(X) \right)\psi_{\tb}^{\mb i}(x)   \d_{i\jb}\d_{\Mb N} \,(\star\, 1)\\[3pt]
&=~\left( \widehat\cD_e D^{\t M}(X) \right)\; \left(\widehat\cD^e D^{\sb \Nb}(X)\right)\, \left(\psi_{\s}^{\,\jb}(x) \star \psi_{\tb}^{\, i}(x)    \right) \,\d_{i\jb}\d_{M \Nb} ~,\label{eq:YMRed2}
\end{split}
\eeq
 where indices for the representation $\rep{R}$ and $\rep{r}$ are explicit. The trace projects onto invariants constructed by the Kr\"onecker delta functions $\d_{i\jb}$ and $\d_{M\Nb}$. In the following we will  suppress the indices and delta symbols where confusion will not arise.  
 
Substituting \eqref{eq:YMRed1} and \eqref{eq:YMRed2} into $\cL_F$ in \eqref{eq:YM}, reintroducing the moduli contribution, calculated in \cite{Candelas:2016usb}, we find a kinetic term for both the matter fields and the moduli fields:
\beq
\begin{split}
 \cL_{F} ~&=~ -2G_{\a\bb} \del_e Y^\a \,\del^e Y^\bb - 2G_{\x\eb} \, \widehat\cD_e C^{\x}\, \widehat\cD^e C^{\eb} -  2G_{\s\tb} \,  \widehat\cD_e D^{\tb}\,  \widehat\cD^e D^{\s}~,
\end{split}\label{eq:matterkinetic}
\eeq
from which we may identify the moduli space metric and matter field metric
\beq
\dd s^2 ~=~ 2G_{\a\bb} \dd y^\a \,\dd y^\bb + 2G_{\x\eb}  \, \dd C^{\x}\dd C^{\eb} + 2G_{\s\tb} \,  \dd D^{\s}\,  \dd D^{\tb}~,
\eeq
where we denote the coordinates of the moduli space $\cM$ by $y^\a, y^\bb$, and without wanting to clutter formulae, denote the coordinates of the matter fields by $C^\x, D^\s$ --- any ambiguity with their corresponding fields will always be made explicit. The moduli fields have the metric computed in \cite{Candelas:2016usb}, given in \eqref{eq:ModuliMetricIntro}. The matter fields also have a metric
\beq
G_{\x\eb} = -\frac{\ap}{4V} \int_\ccX\phi_\eb \star \phi_\x\,  ~, \qquad G_{\s\tb} = -\frac{\ap}{4V} \int_\ccX \psi_\s \star \psi_\tb \,~.\label{eq:mattermetric1}
\eeq
There is no trace as the integrands are written in the form $\brep{r}\cdot\rep{r}$. Although we have indicated the result for a single  representation $\rep{r}$ and $\brep{r}$  the result  generalises to a sum over representations of $\Lh$. 


For any two form $\cF$ there is a relation  
$
\o\star \cF = \half \cF \o^2~.
$
Using this and $\o^{\m\nb} = -\ii g^{\m\nb}$   we find
\beq
\begin{split}
\,\phi_\eb \star \phi_\x  ~&=-\ii \o\star \phi_\x^i \, \phi^\jb_\eb\, \d_{i\jb}  ~=~- \frac{\ii}{2} \o^2 \,\tr (\phi_\x  \,\phi_\eb)~,\cr
  \,\psi_\s \star \psi_\tb  ~&=-\ii \o\star \psi_\s^\jb \, \psi^i_\tb\, \d_{i\jb}~=~ -\frac{\ii}{2}  \o^2 \,\tr (\psi_\s  \,\psi_\tb)~.
\end{split}
\eeq
We introduce the trace over $\rep{r}$ indices in order to be able to write $\phi_x, \psi_\s$ in any order. The matter field metrics are then  expressible in a way closely resembling  the moduli metric
\beq
\begin{split}
G_{\x\eb} ~&=~ \frac{\ii\ap}{8V} \int_\ccX \o^2 \,\tr\,\phi_\x  \, \phi_\eb\, ~, \qquad  \phi_\x\in H^{1}(\ccX,\ccE_{\rep{r}})~,\\
G_{\t\sb} ~&=~  \frac{\ii\ap}{8V} \int_\ccX \o^2 \,\tr \, \psi_\s  \,\psi_\tb~,\qquad \psi_\s\in H^{1}(\ccX,\ccE_{\brep{r}})~.\\
\end{split}\label{eq:mattermetric}
\eeq

\newpage
\section{Fermions and Yukawa couplings}
\label{s4}
The fermionic couplings of interest to heterotic geometry derive from the kinetic term for the gaugino. We compute the quadratic and cubic fluctuation terms. The former are mass terms for the gauginos, which we show all vanish consistent with the vacuum being supersymmetric. The latter  are the Yukawa couplings between two gaugino's and a gauge boson.

In Appendix \sref{app:Spinors} all spinor conventions we used are explained. We also give a summary of results in spinors in $d=4,6,10$ relevant to this section. We also derive some expressions for bilinears relevant to the dimensional reduction. 

\subsection{Fermion zero-modes on $\IR^{3,1}\times \ccX$}

\subsubsection{$\so(3,1)\oplus \su(3)$ spinors}
The gaugino is a  Majorana--Weyl spinor $\ve$ which has zero modes on $\IR^{3,1}\times \ccX$.  The Lorentz algebra is $\so(3,1)\oplus \so(6) \subset \so(9,1)$ under which 
  $\ve$ is
 $$
 \ve ~=~ \z' \otimes \l'\oplus \z\otimes \l  ~\cong~ 
\begin{pmatrix}
 \z'\\
 0
\end{pmatrix}\otimes \l'
+
\begin{pmatrix}
 0\\
 \zeb
\end{pmatrix}\otimes \l ~,
 $$
where $\l$ and $\l'$ are  in the $\rep{4}$ and $\rep{4}'$ of $\so(6)$;  $\z$ and $\z'$ are in the $\rep{2}$ and $\rep{2}'$ of $\so(3,1)$. Where possible we use the $2$-component Weyl notation for $\z,\z'$, and  always leave the $\so(6)$ spinor indices implicit. We write $\oplus$ in this context to reflect the embedding of 2-component spinors into a 4-component notation as shown by the second equality. The barring of 2-component spinors, and dotting the spinor index, comes with complex conjugation $(\z^a)^* = \zeb^\adot$ as described in the appendix. The  fermions $\z,\z'$ are Grassmann odd, and under complex conjugation are interchanged without paying the price of a sign.

The Majorana condition implies $\z\otimes \l$ are determined in terms of $\z'\otimes\l'$. With our conventions this means
\beq\label{eq:so91Majorana2text}
\zeb^\adot \otimes \l ~=~ \zeb'{}^\adot \otimes \l'^c~,
\eeq
where $\l'^c$ denotes taking the $\so(6)$ Majorana conjugate. 
The Majorana-Weyl spinor $\ve$ can now be written solely in terms of say $\z',\l'$:
\beq\label{eq:MajoWeyl}
 \ve ~=~
\begin{pmatrix}
 \z_a'\\
 0
\end{pmatrix}\otimes \l'~
+
\begin{pmatrix}
 0\\
 \zeb'^\adot
\end{pmatrix}\otimes \l'{}^c ~.
\eeq
The presence of $\cN=1$ spacetime supersymmetry means there is a globally well-defined spinor on $\ccX$. This implies the existence of an $\su(3)$--structure on $\ccX$. Under $\so(6) \to \su(3)$
the spinors $\l,\l'$ decompose according to the branching rule   $\rep{4} = \rep{3} \oplus \rep{1}$, and $ \rep{4}' = \brep{3} \oplus \rep{1}$,  which we write as
$$
\l =  \l_{\rep{3}} \oplus  \l_+~,\qquad\qquad \l' =  \l_{\brep{3}} \oplus  \l_-~.
$$
The spinors $\l_+, \l_-$ are the nowhere vanishing $\su(3)$ invariant spinors. As established in the appendix, we can express $\l_{\rep{3}}, \l_{\brep{3}}$ in terms of $\l_\pm$ and gamma matrices
\beq\label{eq:L}
\l_{\rep{3}} ~=~ \L_\m \g^\m \l_-~, \qquad \l_{\brep{3}} ~=~ \L'_\mb \g^\mb \l_+~,
\eeq
where $\{\g^\m , \g^\nb\} = g^{\m\nb}$ and $\L_\m, \L'_\mb$ are components of 1-forms on $\ccX$. The appendix  details the construction of the $\su(3)$ bilinears:
\beq
\label{eq:MetricBilinearText}
\l_+^\dag \g^\m\g^\nb \l_+ ~=~ g^{\m\nb}~, \qquad\l_-^\dag \g^\nb\g^\m \l_- ~=~ g^{\m\nb}~,
\eeq
and
\beq
\label{eq:OmegaSpinorText}
  \O_{\m\n\r}~=~ - e^{-\ii\phi}\,||\O||\, \l_-^\dag \g_{\m\n\r} \l_+~,\qquad  \Ob_{\overline{\m\n\r}}~=~  e^{\ii\phi}\,||\O||\, \l_+^\dag \g_{\ol{\m\n\r}} \l_-~,
\eeq
where $\phi$ accounts for a relative phase difference between $\l_-$ and $\O$. Under the gauge symmetry $\O \to \mu \O$, the fermions $\l_\pm$  transform as a phase:
$$
\l_\pm \to e^{\pm\ii \x/2} \l_\pm~, \quad \mu = |\mu| e^{\ii \x}~.
$$
The bilinears above respect this gauge symmetry. 

Given \eqref{eq:L}, the action of Majorana conjugation is
\beq
\label{eq:lambdaconj2}
\begin{split}
 \l_\pm^c ~&= -\ii  \l_\mp~, \qquad \l_{\brep{3}}^c ~=~  \ii \L'{}_\m^\dag \g^\m \l_-~, \qquad \l_{\rep{3}}^c ~=~ \ii  \L_\mb^\dag \g^\mb \l_+~.
\end{split}
\eeq

\subsubsection{Kaluza Klein ansatz}
$\ve$ is in the adjoint of $\Le_8$. Consequently it decomposes under $\Lg \oplus \Lh \subset \Le_8$ and the expectation from supersymmetry is that we find a natural pairing between fluctuations of the gauge field, and the fermions. As the background is bosonic all fermionic fields are fluctuations, we aim to study the effective field theory of those fluctuations that are massless.  The massless fluctuations are zero-modes of an appropriate Dirac operator. 

The gaugino and gauge field have decomposition under $\so(3,1) \oplus \su(3) \subset \so(9,1)$
\beq
\begin{split}
 \ve:\quad \rep{16} ~&=~ \left(\rep{2} \otimes \rep{1} \oplus \rep{2}'\otimes \rep{1}\right)\oplus \left(\rep{2}\otimes \rep{3} \right)\oplus\left( \rep{2}'\otimes \brep{3}\right)~,\cr
A:\quad \rep{10} ~&=~ \rep{4} \oplus \rep{3} \oplus \brep{3}~,
\end{split}
\eeq
and for the gauge algebra $\Le_8 \supset \Lg\oplus \Lh$:
\beq
{\rm ad}_{\Le_8} ~=~ (\rep{1}, {\rm ad}_{\Lh}) \oplus ({\rm ad}_{\Lg}, \rep{1}) \oplus_i (\brep{R_i}, \rep{r_i}) \oplus_i(\rep{R_i}, \brep{r_i})~.\notag
\eeq 

 We  organise our study of the zero-modes according to their representations under $\so(3,1)\oplus \su(3)$ and the gauge algebra $\Lg \oplus \Lh$.  We continue the mnemonic  of indicating the gauge structure through block matrices
\beq
\begin{split}
 \d A ~&=~ \d A_{(\ad_\Lg,\rep{1})} \oplus_i  \d A_{(\rep{R_i},\brep{r_i})}\oplus_j  \d A_{(\brep{R_j},\rep{r_j})} \oplus \d A_{(\rep{1}, \ad_\Lh)} \\
 &=~ 
\begin{pmatrix}
 \d A_{\ad_\Lh} &\oplus_j \d A_{(\brep{R_j},\rep{r_j})} \\
  \oplus_i\d A_{(\rep{R_i},\brep{r_i})} & \d A_{\ad_{\Lg}}
\end{pmatrix} = 
\begin{pmatrix}
\d \A - \d \A^\dag & \Phi \\
\Psi & \d B
\end{pmatrix}~, \cr
 \ve &=~ \ve_{(\ad_\Lg,\rep{1})} \oplus_i  \ve_{(\rep{R_i},\brep{r_j})}\oplus_j  \ve_{(\brep{R_j},\rep{r_j})} \oplus \ve_{(\rep{1}, \ad_\Lh)}\\
&=~
\begin{pmatrix}
( \z'\otimes \l' \oplus \z'{}^c\otimes \l'{}^c)_{(\rep{1},\ad_\Lh)} & \oplus_j( \z'\otimes \l' \oplus \z'{}^c\otimes \l'{}^c)_{(\brep{R_j},\rep{r_j})} \\
\oplus_i ( \z'\otimes \l' \oplus \z'{}^c\otimes \l'{}^c)_{(\rep{R_i},\brep{r_i})} & (\z'\otimes \l' \oplus \z'{}^c\otimes \l'{}^c)_{(\ad_\Lg,\rep{1})}
\end{pmatrix} ~.
\end{split}\label{eq:gaugestructure}
\eeq
In the second line we have indicated the representations of the individual components of the gaugino by a subscript, and these are regarded as independent field fluctuations. We will now drop the sum over representations $\oplus_i$ in order to simplify notation and as the generalisation to include a sum over representations is obvious. 
%
 
 We classify the zero modes by the symmetries under $\su(3)$-structure and $\Lg \oplus \Lh$. 
 The first type of zero-modes are $\su(3)$-structure singlets, and transform in the  $(\ad_\Lg,\rep{1})$ with  KK ansatz
\beq
\begin{split}
 \d A_{\Le_8\,e}\, \dd X^e ~&=~ 
\begin{pmatrix}
 0 & 0 \\
 0 & \d B_e \dd X^e
\end{pmatrix}~,\cr
\ve_{ (\ad_\Lg,\rep{1})} ~&=~ 
\begin{pmatrix}
 0 & 0\\
 0 & \z_{\ad_\Lg} 
\end{pmatrix} \otimes \l_- \oplus 
\begin{pmatrix}
 0 & 0\\
 0 & \ii\zeb_{\ad_\Lg}  
\end{pmatrix} \otimes \l_+~,
\end{split}\label{eq:fermionsadg}~
\eeq
where we use the first line of \eqref{eq:SpinorBilinearAdG}. 


 The second type of zero-modes transform as $\brep{3}$ under $\su(3)$--structure and as $ (\rep{1},\ad_{\Lh}) \oplus (\rep{R},\brep{r}) \oplus (\brep{R},\rep{r})$   under $\Lg\oplus \Lh$. There is a natural pairing between $\d A^{0,1}_{\Le_8}$ and $\z'\otimes \l_{\brep{3}}$. Using the KK ansatz \eqref{eq:Afluct}, \eqref{eq:MatterKK} for bosons  there is a corresponding KK ansatz for the spinors:
 \beq
\begin{split}
 \d A_{\Le_8}^{0,1} ~&=~ 
\begin{pmatrix}
  Y^\a \,\ccD_\a \A_\mb  & C^{\,\x}\, \phi_{\x\,\mb} \\
  \psi_{\t\,\mb}\,D^{\t} & 0
\end{pmatrix}\,\dd x^\mb~,\\[5pt]
\z'\otimes \l_{\brep{3}} 
~&=~
 \begin{pmatrix}
\cY^\a \, \ccD_\a \A_\mb  &  \cC^{\x}  \,\phi_{\x\,\mb}\,\\
  \cD^{\t} \,  \psi_{\t\,\mb} & 0
\end{pmatrix}\,\otimes\g^\mb \l_+~,
\end{split}\label{eq:fermionzeromodes1}
\eeq
where the matrices in the last line are related to the $\L_\mb'$ in \eqref{eq:L}. 
We have denoted the anticommuting $\IR^{3,1}$ spinors  by   calligraphic letters $\cY^\a$, $\cC^{\x}$ and $\cD^{\t}$, superpartners to $ Y^\a, C^\x, D^\t$.   $\cC^{\x}$ and $\cD^{\t}$ are  in the $\brep{R}$ and $\rep{R}$ of $\Lg$ while $\cY^\a$ are  neutral. 

The spinor in the $\rep{3}$ of $\su(3)$-structure is determined by  Majorana conjugation $\zeb^\adot \otimes \l_{\rep{3}} = \zeb'^\dag{}^\adot\otimes \l_{\brep{3}}^c$, expressed through \eqref{eq:so91Majorana2text} and \eqref{eq:lambdaconj2}: 
\beq
\begin{split}
 \d A_{\Le_8}^{1,0} ~&= -
\begin{pmatrix}
 \d y^\ab\, \ccD_\ab \A^\dag &  D^{\tb}\, \psi_{\tb\,\m}  \\
C^{\xb}\,  \phi_{\xb\,\m} & 0
\end{pmatrix}\,
\dd x^\m~,\\[5pt]
\zeb\otimes \l_{\rep{3}} 
~&=~  \ii
 \begin{pmatrix}
\cYb^\ab \, \ccD_\ab \A^\dag_\m  &  \cDb^{\tb} \, \, \psi_{\tb\,\m} \\
\cCb^{\eb}  \,\,\phi_{\eb\,\m}  & 0
\end{pmatrix}\,\otimes\g^\m \l_-~.
\end{split}\label{eq:fermionzeromodes2}
\eeq
Altogether, the Majorana--Weyl spinor is  given by substituting the above two expressions into  the first line of \eqref{eq:SpinorBilinears1} and combining with \eqref{eq:fermionsadg} giving
\beq
\label{eq:MajoWeyl}
\begin{split}
  \ve ~&=~
 \begin{pmatrix}
 0 & 0\\
 0 & \z_{\ad_\Lg} 
\end{pmatrix} \otimes \l_- \oplus 
\begin{pmatrix}
 0 & 0\\
 0 & \ii\zeb_{\ad_\Lg}  
\end{pmatrix} \otimes \l_+  +  \\
&\quad + 
 \begin{pmatrix}
\cY^\a \, \ccD_\a \A_\mb  &  \cC^{\x}  \,\phi_{\x\,\mb}\,\\
  \cD^{\t} \,  \psi_{\t\,\mb} & 0
\end{pmatrix}\,\otimes\g^\mb \l_+
\oplus \ii
 \begin{pmatrix}
\cYb^\ab \, \ccD_\ab \A^\dag_\m  &  \cDb^{\tb} \, \, \psi_{\tb\,\m} \\
\cCb^{\eb}  \,\,\phi_{\eb\,\m}  & 0
\end{pmatrix}\,\otimes\g^\m \l_-~.
\end{split}
\eeq
The $\oplus$ reflects the embedding of the 2-component Weyl spinors into a 4-component notation used in \eqref{eq:SpinorBilinearAdG}, \eqref{eq:SpinorBilinears1}.

\subsection{Dimensional reduction of $\tr \veb \,\G^M \,D_M \ve$ }
The ten-dimensional kinetic term for the gaugino is now dimensionally reduced, with action:~
\beq
\cL_\ve ~= \frac{ \ap}{4V} \int_\ccX \vol \ii \tr \left(\veb \,\G^M \,D_M \ve\right) ~.
\eeq
 The quadratic fluctuations give the kinetic terms as well as any mass terms; the cubic fluctuations give Yukawa interactions. 

We split the bilinear into two terms
\beq\label{eq:spinorbilinear0}
\tr \left(\veb \,\G^M \,D_M \ve\right) ~=~ \tr \left(\veb \, \G^M\,\del_M \ve\right)\,+\,\tr \left(\veb \, \G^M\,[A_M, \ve]\right)~,
\eeq
where $\veb = \ve^\dag \,\G^0$ is the Dirac conjugate. $\G^M$ are the $d=10$ gamma matrices, given as 
\beq\notag
\G^e = \g^e\otimes \g^{(6)}, \qquad \G^\m = 1\otimes \g^\m~,
\eeq 
as described in the appendix. Here $\m$ is a holomorphic index along $\ccX$. 
\subsubsection{Quadratic couplings}
As the background is bosonic, we take  $A$ to be the background gauge field
\beq\label{eq:backgroundgauge}
A ~=~ 
\begin{pmatrix}
 \A - \A^\dag & 0 \\
 0 & B_e \dd X^e
\end{pmatrix}~.
\eeq

We start with the derivative operator 
 $$
 \ii\tr \veb \, \G^M\,\del_M \ve ~=~\ii \tr \left(\veb \, \G^e\,\del_e \ve\right) +\ii \tr\left( \veb \, \G^m\,\del_m\ve\right)~.
 $$
The first term $\tr \left(\veb \, \G^e\,\del_e \ve\right)$ is computed using \eqref{eq:MajoWeyl} and the third lines of \eqref{eq:SpinorBilinearAdG}, \eqref{eq:SpinorBilinears1},
\beq
\begin{split}
 \vol\ii \tr &\, \Big(\veb \,\G^e \,\del_e \ve \Big)
~=   -2\ii\, \, \vol \tr_\Lg\Big( \zeb_{\ad_\Lg}'\, \sb^{e} \,\del_e\, \z_{\ad_\Lg}' \Big)\, 
+  \Big( \,\cYb^\bb \, \sb^{e} \,\del_e\, \cY^\a \Big)\, \o^2 \tr_\Lh ( \cD_\a \A \cD_\bb \A^\dag)+\\
 &\qquad+   \Big( \,\cC^\x \, \s^{e} \,\del_e\, \cCb^\eb \Big)\, \o^2 \tr_\Lh ( \phi_\x \phi_\eb
)  +   \Big( \,\cDb^\sb \, \sb^{e} \,\del_e\, \cD^\t \Big)\, \o^2  ( \psi_\t \psi_\sb )~.  
\end{split}\label{eq:bilinear0a} \raisetag{.8cm}
\eeq
Spinor and representation indices are contracted in the natural way. 

The second term $\tr\left( \veb \, \G^m\,\del_m\ve\right)$ follows from the fourth lines of  \eqref{eq:SpinorBilinearAdG}, \eqref{eq:SpinorBilinears1} together with the relation in \eqref{eq:wedgeRelation}:
\beq
\begin{split}
\vol& \,\ii  \tr \left(\veb \G^m \del_m \ve\right) 
= -\ii ( \cY^\a\cY^\b )\,\O\, \tr \Big((\cD_\a \A )\,(\delb\, \cD_\b \A)\Big) \frac{e^{-\ii\phi}}{\ON} \\
&\quad+\ii ( \cY^\ab\cY^\bb) \,\Ob\, \tr\Big( (\cD_\ab \A^\dag) \,(\del\, \cD_\bb \A^\dag)\Big) \frac{e^{\ii\phi}}{\ON}  -\ii ( \cC^\x\cD^\tau )\,\O\,\Big( \psi_\tau \,(\delb\, \phi_\x)+ ( \delb\, \psi_\tau)\,\phi_\x  \Big) \frac{e^{-\ii\phi}}{\ON} \\[3pt]
&\quad+\ii ( \cCb^\xb\cDb^\tb) \,\Ob\, \Big(   (\del \phi_\xb^\dag) \psi^\dag_\tb + \phi_\xb^\dag(\del \psi^\dag_\tb ) \Big)\,  \frac{e^{\ii\phi}}{\ON} ~.
\end{split}\label{eq:bilinear0a}\raisetag{.8cm}
\eeq 

Next we compute the reduction of 
\beq\label{eq:spinorbilinear}
\ii\tr \left(\veb \, \G^M\,[A_M, \ve]\right) ~=~ \ii\tr \left(\veb \, \G^e\,[A_e, \ve]\right) + \tr \left(\veb \, \G^m\,[A_m, \ve]\right)~.
\eeq

The first terms follows the calculation of  \eqref{eq:bilinear0a} after using \eqref{eq:backgroundgauge} and $\tr B_e = 0$
\beq
\begin{split}
 \vol\, \ii \tr &\, \Big(\veb \,\G^e \,[A_e, \ve] \Big)
~=~   -2\ii\, \, \vol \tr_\Lg\Big( \zeb_{\ad_\Lg}'\, \sb^{e} \,[B_e, \z_{\ad_\Lg}'] \Big)\, 
+\\
 &\qquad+   \Big( \,\cC^\x \, \s^{e} \,B_e\, \cCb^\eb \Big)\, \o^2 \tr_\Lh ( \phi_\x \phi_\eb
)  +   \Big( \,\cDb^\sb \, \sb^{e} \,B_e\, \cD^\t \Big)\, \o^2  ( \psi_\t \psi_\sb )~.  
\end{split}\label{eq:bilinear1} \raisetag{.5cm}
\eeq

The  second term $\tr \veb \G^m\, [A_m , \ve]$ mirror the calculation of \eqref{eq:bilinear0b}, using the background \eqref{eq:backgroundgauge}
\beq
\begin{split}
\vol& \,\ii  \tr \left(\veb \G^m [A_m ,\ve]\,\right) 
= -\ii ( \cY^\a\cY^\b )\,\O\, \tr \Big((\cD_\a \A) \,\{\A, \cD_\b \A\}\Big) \frac{e^{-\ii\phi}}{\ON} \\[2pt]
&\quad +\ii ( \cY^\ab\cY^\bb) \,\Ob\, \tr \Big((\cD_\ab \A^\dag) \,\{\A^\dag, \cD_\bb \A^\dag\}\Big) \frac{e^{\ii\phi}}{\ON} \\[3pt]
&\quad  -2\ii ( \cC^\x\cD^\tau )\,\O\,\Big( \psi_\tau \,\A\, \phi_\x  \Big) \frac{e^{-\ii\phi}}{\ON}   +2\ii ( \cCb^\xb\cDb^\tb )\,\Ob\,\Big( \phi^\dag_\xb \,\A^\dag\, \psi^\dag_\tb  \Big) \frac{e^{\ii\phi}}{\ON} ~.
\end{split}\label{eq:bilinear0b}\raisetag{1.3cm}
\eeq

We now put the  terms together. The  $~\Dslash_4$ term comes from adding  \eqref{eq:bilinear0a} and \eqref{eq:bilinear1}:
\beq
\begin{split}
& \frac{\ap}{4V} \int_\ccX \vol   \ii\, \tr \left(\veb \,\Dslash_4 \,\ve\right) ~= 
   -\frac{\ii\ap}{2} \tr_\Lg\Big( \zeb_{\ad_\Lg}'\, \sb^{e} \,\hat D_e\, \z_{\ad_\Lg}' \Big)\, 
-2 G_{\a\bb} \,\ii\, \cYb^\bb \, \sb^{e} \,\del_e\, \cY^\a +\\
 &\qquad -2 G_{\x\tb} \,\ii\, \cC^\x \, \s^{e} \,\hat D_e\, \cCb^\eb    -2G_{\t\sb} \,\ii\,\cDb^\sb \, \sb^{e} \,\hat D_e\, \cD^\t ~.  
   \end{split}
\eeq
where $\hat D_e$  is a spacetime covariant derivative appropriate to whatever representation if it acts on 
$$
\hat D_e \z_\Lg ~=~ \del_e \z_\Lg + [B_e,\z_\Lg]~,~~~~~ \hat D_e \cC^\x ~=~ \del_e \cC^\x +  \cC^\x B_e,~~~~~ \hat D_e \, \cD^\t ~=~ \del_e \cD^\t +  B_e\, \cD^\t~.$$
The matter and moduli fields have kinetic terms with non-trivial metrics:
 \beq
\begin{split}
G_{\a\bb} ~&=~  
 \frac{\ii\ap}{8V} \int_\ccX \o^2 \,\tr_\Lh\,  \ccD_\a \A \ccD_\bb \A^\dag~,\\[5pt]
G_{\t\sb} ~&=~  
 \frac{\ii\ap}{8V} \int_\ccX \o^2 \,\psi_\t  \,\psi_\sb~,\qquad
G_{\x\eb} ~=~ 
\frac{\ii\ap}{8V} \int_\ccX \o^2 \,\tr_{\Lg}\,   \phi_\x\,\phi_\eb\, ~.\\[5pt]
\end{split}\label{eq:mattermetrics}
\eeq
The fermions have identical metrics to their bosonic superpartners.  The bundle moduli appear with a metric that coincides with that derived by Kobayashi and Itoh \cite{Kobayashi:1987, Itoh:1988}. 

The mass terms come from adding together \eqref{eq:bilinear0a} and \eqref{eq:bilinear0b}. We normalise the mass term to be compatible with the convention in \cite{WessBagger}:
\beq
\begin{split}
  \frac{ \ap}{4V}& \int_\ccX \vol  \ii \tr \left(\veb \,\G^m D_m  \,\ve\right) ~= - e^{\ccK/2} m_{\a\b} \, ( \cY^\a\cY^\b)   -2e^{\ccK/2} m_{\x\t}\, ( \cC^\x \cD^\tau) + {\rm c.c.}
\end{split}
\eeq
where $\ccK$ is the \K potential, which on our background evaluates to be $e^{\ccK/2} = (2\sqrt{2} V \ON)^{-1/2}$. The mass terms are  
\beq
\begin{split}
 m_{\a\b} ~&=~\frac{\ii \ap e^{-\ii \phi}}{\sqrt{2}} \int_\ccX  \O\, \tr_{\Lh} \Big(\ccD_\a \A \,(\delb_\A \, \ccD_\b\A) \,\Big)~,
 \\[5pt]\label{eq:mass0}
 m_{\x\tau} ~&=~\frac{\ii \ap e^{-\ii \phi}}{2\sqrt{2}} \int_\ccX  \O\, \, \Big(\psi_\tau \,(\delb_\A \, \phi_\tau) + (\delb_\A \psi_\tau) \, \phi_\tau ) \,\Big)~.
 \\[5pt]
 \end{split}
\eeq
The last term is normalised with a factor of $2$ as the two indices are distinguished. 
As before, we do not write the trace, understanding the indices contracted in the natural way. 

Recall that the equations of motion are
\beq
\begin{split}
 \delb_\A (\ccD_\a\A) ~&=~ \D_\a{}^\m F_\m~,\quad  \delb_\A \phi_\x  ~=~ 0~, \qquad \delb_\A \psi_\t  ~=~  0~.
\end{split}
\eeq
Substituting this in we find
\beq
\begin{split}
  m_{\a\b} ~&=~-\frac{\ii \ap e^{-\ii \phi}}{4V\ON} \int_\ccX  \O\, \D_\b{}^\m \tr_{\Lh} \Big(\ccD_\a \A \, F_\m\,\Big)~,\\
m_{\x\t} ~&=~ 0~.
\end{split}
\eeq
The vanishing of $m_{\a\b}$ is guaranteed when $\delb( \ccD_\a \A )= 0$, that is for bundle or hermitian moduli.  For complex structure parameters, if one can find a basis for parameters in which $\delb(\ccD_\a \A) = 0$ for complex structure moduli, so that $\D_\a{}^\m F_\m = 0$, then this is also satisfied. If this is not possible we exploit the last line of the supersymmetry relation \eqref{eq:cBrelations1}
\beq
\begin{split}
 \frac{\ii\ap}{4} \int_\ccX  \O\, \D_\b{}^\m \tr_{\Lh} \Big(\ccD_\a \A \, F_\m\,\Big) ~&=~ \int_\ccX \O\, \D_\a{}^\m \D_\b{}^\n  (\del_\m \o_\n^{0,1} - \del_\n \o_\m^{0,1})\\
 &=~ \int_{\ccX} (\D_\a{}^\m \D_\b{}^\n \O_{\m\n}^{1,0})\,\del\o \\ 
 &=-\ii e^{-\ccK_2} \mathfrak{a}_{\a\b}{}^\gb\int_{\ccX}  \ol{\chi}_\gb\ \del \o \\
 &=~ 0~.
\end{split}\label{eq:vanishingmass}
\eeq
We have used some results familiar from special geometry, see \cite{Candelas:1990pi}, which apply in this general heterotic context. They are:
$$
(\D_\a{}^\m \D_\b{}^\n \O_{\m\n}^{1,0}) = (\cD_\a \chi_\b)^{1,2} = - \ii e^{\cK_2} \mathfrak{a}_{\a\b}{}^\gb \ol{\chi}_\gb~,\quad \cD_\a \chi_\b ~=~ \del_\a \chi_\b + (\del_\a \ccK_{2}) \chi_\b~,
$$
Here $\cD_\a \chi_\b$ is covariant with respect to gauge transformations $\chi_\a \to \mu\, \chi_\a$. Also,
$$
\mathfrak{a}_{\a\b\g} ~=~ - \int \O \D_\a{}^\m \D_\b{}^\n \D_\g{}^\r \ \O_{\m\n\r}~, \qquad \mathfrak{a}_{\a\b}{}^\gb ~=~ \mathfrak{a}_{\a\b\g} G_{0}{}^{\g\gb}~, \quad G_{0\,\a\bb} ~=-\frac{\int \chi_\a\chi_\bb}{\int \O \Ob}~.
$$
and 
$$
\ccK_2 ~=~ - \log \Big(\ii \int \O \, \Ob \Big)
$$
In special geometry $\mathfrak{a}_{\a\b\g}$ is a Yukawa coupling for $\rep{27}^3$ fields,  playing  the role of an intersection quantity relating derivatives of $\chi_\a$ to $\ol{\chi}_\gb$.  $G_0$ the metric on complex structures, used to raise and lower indices. The same relation applies in heterotic geometry with the understanding the complex structures may be reduced by the Atiyah constraint. Consequently, we see the Atiyah condition gives
$
m_{\a\b} ~=~ 0~.
$

\vskip-15pt
\subsubsection{Cubic fluctuations and Yukawa couplings}
We now compute the cubic order  fluctuations to get the Yukawa couplings. The calculation proceeds in a similar fashion to the above. The cubic interaction only come from:
\beq\label{eq:cubic_bilinear}
\tr \left(\veb \, \G^M\,[\d A_M, \ve]\right)~.
\eeq
The fluctuations only occur on the internal space $\d A_M \G^M = \d A_m \G^m$. The gauge structure of $\d A$ is specified in \eqref{eq:gaugestructure}.   The calculation is a simple generalisation of the result \eqref{eq:bilinear0b} using the fourth line of \eqref{eq:SpinorBilinears1}. 
\beq
\begin{split}
\vol& \,\ii  \tr \left(\veb \G^m [\d A_m ,\ve]\,\right) 
= -\ii ( \cY^\a Y^\b \cY^\g )\,\O\, \tr \Big((\cD_\a \A) \,\{\cD_\b\A, \cD_\g \A\}\Big) \frac{e^{-\ii\phi}}{\ON} \\[2pt]
&\quad -2\ii \Big( \cC^\x Y^\a \cD^\t +  \cC^\x \cY^\a D^\t  +  C^\x \cY^\a \cD^\t \Big)\,\O\, \tr (\psi_\t \,\cD_\a \A\, \phi_\x) \frac{e^{-\ii\phi}}{\ON}  \\[3pt]
&\quad -\ii \tr_\Lg( \cD^\tau D^\s \cD^\r )\,\O\, \tr \Big(\psi_\tau \{\psi_\s, \psi_\r\} \Big) \frac{e^{-\ii\phi}}{\ON} \\
&\quad -\ii \tr_\Lg( \cC^\x C^\eta \cC^\pi )\,\O\, \tr \Big(\phi_\x \{\phi_\eta, \phi_\pi\} \Big) \frac{e^{-\ii\phi}}{\ON}  +{\rm c.c.}
\end{split}\label{eq:bilinear5}\raisetag{.5cm}
\eeq 

%
The last two lines, we use the appropriate symmetric invariants to constructing $\rep{R}^3$ and $\brep{R}^3$.


Putting it together, normalising to agree with \cite{WessBagger}, we find
\beq
\begin{split}
  \frac{ \ap}{4V}&\int_\ccX  \vol \ii \tr\left( \veb \, \G^M\,[\d A_M, \ve ]\right)~=- 4 e^{\ccK/2} \ccY_{\x\a\t} (\cC^\x Y^\a \cD^\t +  \cC^\x \cY^\a D^\t  +  C^\x \cY^\a \cD^\t)\\[4pt]
 &- 2e^{\ccK/2} \ccY_{\x\eta\pi} (\cCb^\x C^\eta \cC^\pi) - 2e^{\ccK/2} \ccY_{\r\s\t}(\cDb^\r D^\s \cD^\t) - 2e^{\ccK/2}  \ccY_{\a\b\g} \left(\cYb^\a  \, Y^\b\,\cY^\g\, \right)  + {\rm c.c.}~,
\end{split}\label{eq:fermionsResult2}
\eeq
where ${\rm c.c}$ denotes the complex conjugate and the Yukawa couplings are given by
\beq
\begin{split}
 \ccY_{\x\a\t} ~&=~ \frac{\ii\ap e^{-\ii\phi}}{2\sqrt{2}} \int_\ccX \O \tr \Big(  \psi_\t\, \cD_\a \A\, \phi_\x \Big)~,\cr
  \ccY_{\a\b\g} ~&=~ \frac{\ii\ap e^{-\ii\phi}}{2\sqrt{2}} \int_\ccX  \O \tr \Big( (\ccD_\a \A)\{\ccD_\b \A, \ccD_\pi\A\}\big)~,\cr
   \ccY_{\x\eta\pi} ~&=~ \frac{\ii\ap e^{-\ii\phi}}{2\sqrt{2}} \int_\ccX  \O \tr \Big( \phi_\x\{\phi_\eta, \phi_\pi \}\big)~,\qquad \ccY_{\t\s\r} ~=~ \frac{\ii\ap e^{-\ii\phi}}{2\sqrt{2}} \int_\ccX  \O \tr \Big( \psi_\t\{\psi_\s, \psi_\r \}\big)~
    \label{eq:Yukawa}
\end{split}
\raisetag{2cm}
\eeq
 The result is straightforwardly extended to vacua with more complicated branching rules involving multiple representations $\oplus_p \rep{R}_p$. More Yukawa couplings appear -- one for each invariant computed via the trace -- but the integrand is of the same form as above. 
 
 The $\rep{1}^3$ coupling vanishes classically. To see this, write $\d \A$ to second order in deformations
$$
\d \A = \d y^\a \ccD_\a \A + \d y^\a \d y^\b \ccD_{\a} \ccD_{\b} \A + \cdots~.
$$
Note that $\ccD_\a \ccD_\b \A = \ccD_\b \ccD_\a \A$ and so the second term is appropriately symmetric in indices $\a,\b$. 
A standard deformation theory argument related to the Kuranishi map implies the second order deformation is unobstructed provided
$$
\delb_\A (\ccD_{\a} \ccD_{\b} \A) + \{\ccD_{\a} \A ,\ccD_{\b} \A \} ~=~ 0~. 
$$
Substituting into \eqref{eq:Yukawa2} one finds 
$
\ccY_{\a\b\g} = 0~,
$
when $\D_\a{}^\m F_\m = 0$. When $\D_\a \ne 0$, that is complex structure is varying, the coupling still vanishes after using exactly argument provided for the singlet mass term in \eqref{eq:vanishingmass}. 

The coupling $Y_{\x\a\t}$ also vanishes by demanding that $\phi_\x$, equivalently, $\psi_\t$, remain solutions of the equation of motion under a bundle deformation $\A \to \A + \d y^\a \ccD_\a \A$: $\delb_{\A + \d \A} \phi_\x = \delb_{\A + \d \A} \psi_\t = 0$. Hence,  the singlet couplings vanish. 
$$
\ccY_{\x\a\t} = \ccY_{\a\b\g} = 0~.
$$

\newpage
\section{The final result: moduli, matter metrics and Yukawa couplings}
\label{s5}
The effective field theory has $\cN=1$ supersymmetry, with a gravity multiplet and a gauge symmetry $\Lg$. The $\cN=1$ chiral multiplets consist of~\footnote{We do not consider the universal multiplet, the $d=4$ dilaton and $B$-field, which decouples.}
\begin{itemize}
 \item $\Lg$-neutral scalar fields $Y^\a$ and fermions $\cY^\a$  corresponding to moduli;
 \item $\Lg$-charged bosons $C^\x$ and fermions $\cC^\x$ in the $\brep{R}$ of $\Lg$; 
 \item $\Lg$-charged bosons $D^\r$ and fermions  $\cD^\r$ in the $\rep{R}$ of $\Lg$;
\end{itemize}

 The final result is expressed as a Lagrangian with normalisation conventions matching \cite{WessBagger}
\beq
\begin{split}
 \cL ~&= -2G_{\x\eb} \del_e Y^\a \,\del^e Y^\eb - 2G_{\x\eb} \, \widehat\cD_e C^{\x}\, \widehat\cD^e C^{\eb} -  2G_{\s\tb} \,  \widehat\cD_e D^{\tb}\,  \widehat\cD^e D^{\s}  
  -\frac{\ii\ap}{2} \tr_\Lg\Big( \zeb_{\ad_\Lg}'\, \sb^{e} \,\hat D_e\, \z_{\ad_\Lg}' \Big)\, \\[3pt]
&\qquad -2 G_{\a\bb} \,\ii\, \cYb^\bb \, \sb^{e} \,\del_e\, \cY^\a  -2 G_{\x\tb} \,\ii\, \cC^\x \, \s^{e} \,\hat D_e\, \cCb^\eb    -2G_{\t\sb} \,\ii\,\cDb^\sb \, \sb^{e} \,\hat D_e\, \cD^\t \\[5pt] 
   & - \Big( e^{\ccK/2} m_{\a\b} \, ( \cY^\a\cY^\b)   2e^{\ccK/2} m_{\x\t}\, ( \cC^\x \cD^\tau) + {\rm c.c.}  \Big)   - \Big(4 e^{\ccK/2} \ccY_{\x\a\t} (\cC^\x Y^\a \cD^\t +  \cC^\x \cY^\a D^\t  +  C^\x \cY^\a \cD^\t)\\[4pt]
 &+ 2e^{\ccK/2} \ccY_{\x\eta\pi} (\cCb^\x C^\eta \cC^\pi) + 2e^{\ccK/2} \ccY_{\r\s\t}(\cDb^\r D^\s \cD^\t) + 2e^{\ccK/2}  \ccY_{\a\b\g} \left(\cYb^\a  \, Y^\b\,\cY^\g\, \right)  + {\rm c.c.} \Big)~.  
\end{split}\raisetag{0.8cm}
\eeq 
The kinetic terms for fields contain metrics. The metric for fermions and bosons are identical, consistent with supersymmetry. The moduli metric, derived in \cite{Candelas:2016usb}, is:
\beq
\begin{split}\label{eq:ModuliMetricIntro}
 \dd s^2 ~&=~  2G_{\a\bb} \,\dd y^\a  \otimes  \dd y^\bb~, \\[6pt]
G_{\a\bb} ~&=~  \frac{1}{4V} \int \D_\a{}^\m \star \D_\bb{}^\n \,\, g_{\m\nb} + \frac{1}{4V} \int \cZ_\a \star \cZ_\bb \,+ \\[3pt]
&\quad + \frac{ \ap}{4V}\int \tr \Big( D_\a A \star D_\bb A \Big) - \frac{ \ap}{4V}\int \tr \Big(D_\a \Th\,\star\, D_\bb \Th^\dag\Big)~.
\end{split}
\eeq
The metric terms for the fermionic superpartners to moduli $\cY^\a$ are fixed by supersymmetry from the the bosonic result.  The matter field metrics are given in \eqref{eq:mattermetric}, 
\beq
\begin{split}
G_{\x\eb} ~&=~ \frac{\ii\ap}{8V} \int_\ccX \o^2 \,\tr\,\phi_\x  \, \phi_\eb\, ~, \qquad  \phi_\x\in H^{1}(\ccX,\ccE_{\rep{r}})~,\\
G_{\t\sb} ~&=~  \frac{\ii\ap}{8V} \int_\ccX \o^2 \,\tr \, \psi_\s  \,\psi_\tb~,\qquad \psi_\s\in H^{1}(\ccX,\ccE_{\brep{r}})~.\\
\end{split}\label{eq:mattermetricb}
\eeq

 The mass terms written in \eqref{eq:mass0} vanish
$
m_{\a\b} ~=~ m_{\x\t} ~=~ 0~.
$

 The Yukawa non-zero couplings  in \eqref{eq:Yukawa} are
\beq
\begin{split}
   \ccY_{\x\eta\pi} ~&=~ \frac{\ii\ap e^{-\ii\phi}}{2\sqrt{2}} \int_\ccX  \O \tr \Big( \phi_\x\{\phi_\eta, \phi_\pi \}\big)~,\qquad \ccY_{\t\s\r} ~=~ \frac{\ii\ap e^{-\ii\phi}}{2\sqrt{2}} \int_\ccX  \O \tr \Big( \psi_\t\{\psi_\s, \psi_\r \}\big)~.
    \label{eq:Yukawa2}
\end{split}
\raisetag{2cm}
\eeq

\newpage

\section{The superpotential and \K potential }
\label{s6}
The effective field theory has $\cN=1$ supersymmetry in $\IR^{3,1}$, and so the couplings ought to be derivable from a superpotential and \K potential. The \K potential for the moduli metric couplings was proposed in \cite{Candelas:2016usb}, and checked against a dimensional reduction of the $\ap$-corrected supergravity action. It is 
\beq
 \ccK{\,}_{moduli} ~=~  - \log\left( \frac{4}{3} \int \o^3\right)  -\log\left( \ii \!\int\! \O\, \Ob \right)~. \label{eq:KahlerPotential}
\eeq
in which $\o$ is  the hermitian form of $\ccX$. The $\ap$-corrections preserved the form of the special geometry \K potential, and the second term remains classical.  

 The \K potential for the matter field metric is trivial and given by 
\beq
\ccK_{\ matter} ~=~ G_{\x\eb}  C^{\x\,M} C^{\eb \Nb} \d_{M\Nb} + G_{\r\tb} \d_{M\Nb} D^{\t M} D^{\rb\Nb} ~,\label{eq:KahlerMatter}
\eeq
where $a,b=1,\ldots, R$ label the $\rep{R}$ representation and the trace is taken with respect to the delta function. 

The F-term couplings for the $d=4$ chiral multiplets are described by a superpotential. In the language of $d{=}4$ effective field theory, this superpotential takes the general form
\beq
\begin{split}
 \ccW(Y^\a, C^\x, D^\t) ~&=~ \frac{1}{3} \ccY_{\x\eta\pi} \tr C^\x C^\eta C^\pi +  \frac{1}{3}\ccY_{\r\t\s} \tr D^\r D^\t D^\s + \cdots~,
\end{split}\label{eq:superpotential0}
\eeq
where the $\tr$ projects onto the appropriate $\rep{R}$-invariant and we are to view these as chiral multiplets in $N=1$ $d{=}4$ superspace in the usual way. The omitted terms are the quartic and higher order couplings and non-perturbative corrections.  It is important that $\ccW$ gives no singlet couplings, and this means all parameter derivatives of $\ccW$ vanish. 

We would like to study a superpotential in a similar vein to the \K potential proposal \eqref{eq:KahlerPotential}.  As ten-dimensional fields $A_{\Le_8}$ and $H$ depend on both parameters and matter fields. The fields $\ddc \o$ and $\O$ are valued on $\ccX$ and depend only on moduli fields. The spirit of the dimensional reduction is to promote the parameters to $d{=}4$ fields. In this vein  define a superpotential~\footnote{The form of this integrand is due to Xenia de la Ossa who suggested to me in private conversation. }  
\beq
\ccW(Y^\a, C^\x, D^\t) ~=~  -\ii \sqrt{2} e^{-\ii\phi} \int \Omega\Big( H - \dd^c \o\Big)~,\label{eq:Superpotential}
\eeq
in which the fields are regarded as functionals of the $d{=}4$ chiral multiplets.  The couplings in the effective field theory are specified by differentiating $\ccW$   and evaluating the integral after fixing the parameters $y=y_0$.   

The rules for differentiating fields in the expressions for  $\ccK$ and $\ccW$ with respect to parameters have been described in \cite{Candelas:2016usb}, which is complicated by virtue of $\Lh$ gauge transformations being parameter and coordinate dependant. These transformations are, however, independent of matter fields, and so the rule for  matter field differentiation is simple
$$
\del_\x A_{\Le_8} ~=~ \frac{\del A_{\Le_8}}{\del C^\x}  ~=~ \phi_\x~.
$$
It is important that we have written the ten-dimensional  $\Le_8$ gauge field $A_{\Le_8}$,  and not $A_\Lh$, as this is the functional of the matter fields -- $C^\x, D^\t$ -- as illustrated in, for example  \eqref{eq:Afluct} and \eqref{eq:MatterKK}. The integrand in $\ccW$ is a functional of the ten-dimensional  $H$  so that it depends on matter fields. The rule is to differentiate as noted above, and then evaluate the integral on the fields' vacuum expectation values (VEV). Note that it is the VEV of $H$ that satisfies $\ddc \o = H$, and the matter fields VEVs vanish $C^\x = D^\t = 0$. 

For example, the tadpole matter and moduli couplings  for a vacuum at the point $y=y_0$ are
\beq
\begin{split}
 (\del_\x \ccW)|_{y=y_0}  ~&\sim~   \int  \O\, \del_\x H |_{y=y_0} ~\sim~ \int \O \tr F \phi_\x|_{y=y_0} ~=~ 0~,\\
 (\del_\a \ccW)|_{y=y_0}  ~
 &\sim~  \int \Big(\,(\chi_\a - k_\a \O) (H-\dd^c\o) +\O \left(\delb (\ccB^{0,2}_\a +\ii \ccD_\a \o^{0,2})\right)\Big)|_{y=y_0} ~=~  0~.\\
\end{split}\label{eq:tadpole}
\eeq
where  we use $\del_\a H$ in  \eqref{eq:HDeriv} and $\del_\a \ddc\o$ in \eqref{eq:dcomDeriv}, and  we evaluate them on some fixed $y=y_0$.    

As an ansatz $\ccW$ must satisfy a number of tests: it must be a section of a line bundle over the moduli space; any derivative with respect to parameters must vanish viz. $\del_\a\del_\b \del_\g \cdots \ccW = 0$;  be a holomorphic function of chiral fields; tadpole and mass terms for the matter fields must vanish; capture the F-term couplings derived through dimensional reduction in this paper. The expression \eqref{eq:Superpotential} passes these tests. 
\footnote{In the literature a different ansatz is proposed for the superpotential:
$
\wt \ccW ~=~ \int \O (H+\ii \dd \o)~.
$
After careful calculation one can check  $\del_\a \wt\ccW = \del_\a \del_\b \wt\ccW = \del_\a \del_\b \del_\g \wt\ccW = 0$, and so there are no $\rep{1}, \rep{1}^2, \rep{1}^3$ couplings. To what extent this reproduces singlet couplings to higher order is an interesting question. }
 
$\ccW$ is a section of the line bundle transforming under the gauge symmetry $\O \to \mu(y) \O$ as $\ccW\to\mu\ccW$ where $\mu(y)$ is a holomorphic function of parameters. This necessary in order to consistently couple to gravity \cite{Witten:1982hu}. This fixes the integrand to be proportional to $\O$. 

The supersymmetry relation $H= \ddc\o$ holds for all $y_0\in \cM$. Hence, derivatives of it vanish:\footnote{Many examples of relations involving complex structure do not hold for all $y_0\in \cM$. A simple example is $\dd J$. Although for any fixed complex structure $\dd J|_{y=y_0} = 0$, differentiating we get something non-zero $\del_\a \dd J = \del \D_\a|_{y=y_0} \ne 0$. }
\beq\label{eq:susy1}
\del_{\a_1} \cdots \del_{\a_n} (H-\ddc \o)|_{y=y_0}~=~ 0~, \qquad \del_{\bb_1} \cdots \del_{\bb_n} (H-\ddc \o)|_{y=y_0} ~=~ 0~,
\eeq
where $y^{\a_1}, \cdots, y^{\a_n}$ are any collection of parameters and we evaluate on a supersymmetric vacuum, denoted by $y=y_0$. It then follows that any derivative of the superpotential with respect to parameters vanishes. This is what is used in \eqref{eq:tadpole} to show that all tadpole terms vanish. The argument clearly extends to higher order. Consider the $k$th derivative
\beq
\begin{split}
\Big(  \del_{\a_1}\cdots\del_{\a_k} \ccW\, \Big)|_{y=y_0} ~&=~  \int\Big(  (\del_{\a_1}\cdots\del_{\a_k}  \O )\, (H-\ddc\o) + \\
&\qquad\qquad+ k\del_{\{\a_2}\cdots\del_{\a_k} \O\, \del_{\a_1\}}( H- \ddc\o) + \cdots\Big)|_{y=y_0} ~=~ 0~.
\end{split}\notag
\eeq 
This vanishes on any supersymmetric background: $\ccW$ is independent of moduli fields, and so $\ccW$ does not give rise to any singlet couplings in agreement with the dimensional reduction.

An analogous argument, together with $\O$ being holomorphic, shows that despite neither $H$ nor $\dd^c\o$ being holomorphic, $\ccW$ is a holomorphic function of fields. For example, the first order derivative is
 \beq
 \del_\ab\,\frac{1}{\O_0}\int \Big(\O (H- \dd^c\o)\Big) ~=~\frac{1}{\O_0} \int \delb\Big(\O (\ccB_\ab^{0,2} - \ii \ccD_\ab \o^{0,2})\Big)|_{y=y_0} ~=~ 0.\notag
 \eeq
Using \eqref{eq:susy1} all higher order anti-holomorphic derivatives of $\O(H-\ddc \o)$  vanish. It is also the case that $(\del_\xb)^n \ccW = 0$ for all $n\ge 1$. 
 So, $\ccW$ is a holomorphic function of chiral fields.

 The expression for the masses can be written as derivatives of $W$ 
 \beq
 m_{\a\b} ~=~    \del_\a \del_\b W~=~ 0, \qquad m_{\x\t} ~=~    \del_\x \del_\t W~=~0~,
 \eeq
 where for the second term we use that $ \dd^c \o, \O$ do not depend on $C^\x, D^\t$ while $\del_\x \del_\t H$ is given by \eqref{eq:Hsecondderiv} with $D_a A \to \del_\x A = \phi_\x$. As $A$ depends linearly on the matter fields, all second derivatives vanish.  
  
 The Yukawa couplings $\ccY$ are also all derived from $\ccW$.   Using \eqref{eq:thirdderiv}, we find agreement with the functional forms in \eqref{eq:Yukawa}, of which the non-vanishing terms are 
\beq
\begin{split}
 \ccY_{\x\eta\pi} ~&=~   \half \del_\x \del_\eta \del_\pi \ccW~, \quad \ccY_{\s\t\r} ~=~ \half \del_\x\del_\a\del_\t \ccW~.
\end{split}
\eeq
Even though the singlet couplings vanish, one can check that their functional form is correctly derivable from $\ccW$. The fact of $1/2$ is in order to agree with the convention given in \cite{WessBagger}. 
It is satisfying that the superpotential consistently captures the couplings derived in the dimensional reduction, both involving moduli and matter fields.  Furthermore, it manifestly does not give rise to any singlet couplings.


%
\newpage
\section{Outlook}
We have calculated the effective field theory of heterotic vacua of the form $\IR^{3,1}\times \ccX$ at large radius, correct to order $\ap$. The field theory is  specified by a \K potential and superpotential. 
Supersymmetry forbids $\ccW$ from being corrected perturbatively in $\ap$, but is in general corrected non-perturbatively in $\ap$. For $\ccE$ obtained by deforming $\ccT_\ccX$, some of these non-perturbative corrections have been computed as functions of moduli using linear sigma models, for example \cite{McOrist:2007kp,McOrist:2008ji,McOrist:2011bn, McOrist:2010ae,Melnikov:2012hk}. One can now use the results obtained here and those in \cite{Candelas:2016usb} to determine the normalised quantum corrected Yukawa couplings, in examples that may be of phenomenological interest, for example \cite{Anderson:2013xka}. Although the \K potential is corrected perturbatively in $\ap$, it was conjectured in \cite{Candelas:2016usb} that the form of the \K potential does not change to all orders in perturbation theory, and that the $\ap$-corrections are contained within the hermitian form $\o$. This conjecture is consistent with the work in \cite{delaOssa:2014cia,delaOssa:2014msa} and it would be very interesting to prove this conjecture, at least to second order in $\ap$. 

Although we have derived this result using a single pair of matter fields, the result clearly generalises to a sum over representations $\oplus_p \rep{R}_p \oplus_p \brep{R}_p$. The main burden of the generalisation is to evaluate the trace using the appropriate branching rules. 

Many questions arise. For example,  are there any special geometry type relations between $\ccK$ and $\ccW$? Finding a prepotential analogous to  special geometry looks difficult, partly because it involved analysis related on the geometry of the standard embedding and \cym's. Nonetheless, it is likely $\ccK$ and $\ccW$ are related. 

It would be interesting to compute the field theory couplings in specific examples. For $\ccE$ attained by deforming  $\ccT_\ccX$ one might be able to compare with the linear sigma model parameter space studied in say  \cite{McOrist:2008ji,Melnikov:2010sa,Kreuzer:2010ph} and study the quantum corrections to the $\rep{27}^3$ and $\brep{27}^3$ couplings using the correctly normalised fields. We showed using deformation theory arguments that the $\rep{1}^3$ coupling vanishes classically. A pressing question is to what extent these couplings vanish exactly. Any non-vanishing would the vacuum does not exist, shrinking the moduli space of heterotic vacua.  

\section*{Acknowledgements}
\vskip-10pt
It is a pleasure to thank Philip~Candelas, Emily~Carter and Xenia~de~la~Ossa for many interesting and helpful conversations related to this work.  I would like to acknowledge the hospitality of Mathematical Institute, University of Oxford, where part of this work was completed. I am supported by STFC grant ST/L000490/1. 
\newpage
\appendix

\section{Hodge theory on real and complex manifolds}
\label{app:Conventions}
We establish some notation and results for forms on real and complex manifolds to be used in the text. Coordinates for $\IR^{3,1}$ are denoted $X^e$ while real coordinates on $\ccX$ are denoted by $x^m$. Complex coordinates are denoted $x^\m, x^\nb$.   

We need to write coordinate expressions for forms more than metrics and so our convention is to omit the wedge symbol $\wedge$ except where confusion may arise. We write metrics as $\dd s^2 = g_{mn} \dd x^m \otimes \dd x^n$, only occasionally omitting the $\otimes$ only where confusion will not arise. 

\subsection{Real manifolds}

The volume form on a $n$-dimensional Riemannian manifold is
\beq
\vol ~=~ \star 1 ~=~ \frac{\sqrt{g}}{n!} \,\e_{m_1\ldots m_{n}}\, \dd x^{m_1} \ldots \dd x^{m_{n}}~,\qquad g = |\det g_{mn}|~.\label{eq:RealVol}
\eeq
where $\e_{12\cdots n} = \e^{12\cdots n} =1$ is the permutation symbol. The determinant of the metric is
$$
g ~=~ \frac{1}{n!} \e^{p_1\cdots p_n} \e^{q_1\cdots q_n} g_{p_1q_1}\cdots g_{p_n q_n}~.
$$
If $\o$ is a top-form then
$$
\vol\, \,\frac{\sqrt{g}}{n!} \,\,\e^{m_1\ldots m_{n}} \o_{m_1\ldots m_n} = \frac{1}{n!} \o_{m_1\ldots m_n} \dd x^1 \ldots \dd x^n~.
$$

The Hodge dual of a $p$-form $A_p$  is
$$
\star A_p ~=~ \frac{\sqrt{g}}{p!(n-p)!} \e^{m_1\ldots m_p}{}_{n_1\ldots n_{n-p}} A_{m_1\ldots m_p} dx^{n_1} \ldots dx^{n_{n-p}}~.
$$
The inner product of two $p$-forms is then
$$
A_p \w \star B_p ~=~ \vol\,\left(\frac{1}{p!}\,\, A_{m_1 \ldots m_p}\,B^{m_1\ldots m_p} \right).
$$

\subsection{Complex manifolds}
On a complex manifold the metric is hermitian
$$
\dd s^2 = 2 g_{\m\nb} \,\dd x^\m \otimes \dd x^\nb
$$
with $\det(g_{mn}) {=} g$. In addition to the Hodge dual $\star$, which contracts a $(p,q)$ with a $(q,p)$ form, on a complex manifold we can define a $\starb$  which contracts a pair of $(p,q)$-forms, and so forming an inner product. If $\a,\b$ are two $(p,q)$-forms then it is defined as 
\beq
\a \starb \b ~:=~\frac{1}{p!q!}\, \a_{\m_1\cdots \m_p \nb_1\cdots\nb_q} \b^{\m_1\cdots \m_p \nb_1\cdots\nb_q} (\star 1 ) ~=~ \a \star (\b)^*~. 
\eeq
It has a complex volume form, which is nowhere vanishing and globally well-defined: 
$$
\O = \frac{1}{3!} \O_{\m\n\r} \dd x^\m\dd x^\n \dd x^\r~, \qquad ||\O||^2 = \frac{1}{3!} \O_{\m\n\r} \Ob^{\m\n\r}~,
$$
where $||\O||$ is a coordinate scalar, and so a constant for a fixed manifold, but depends on parameters, denoted $y$. We can write
\beq\label{eq:OmDef}
\O_{\m\n\r} = f(x,y) \e_{\m\n\r}~, \qquad \e_{123} = 1~,
\eeq
where $e_{\m\n\r}$ is the permutation symbol and $f(x,y)$ is a holomorphic function of coordinates and parameters.  $\e_{\m\n\r}$ is not a tensor and consequently $f$ transforms like $g^{1/4}$ under holomorphisms (holomorphic diffeomorphisms): if $x' = x'(x)$ then $f(x)$ transforms
$f'(x) = (\det j) f(x)$, where $j^\m_\n = \pd{x'{}^\m}{x^\n}$.  It satisfies the relation
\beq\label{eq:OmRelation}
|f|^2 = g^{1/2} ||\O||^2~,
\eeq
The complex volume form $\O$ transform like  sections of a complex line bundle on $\ccM$. Under a gauge transformation $\O \to \mu \O$ with $\mu\in \IC^*$,   we have $||\O||^2 \to |\mu|^2 ||\O||^2$ while $g$ is invariant. It is sometimes convenient to isolate the phases of $f$ and $\m$:
$$
f ~=~ |f| e^{\ii \z}~, \qquad \m = |\m| e^{\ii \x}~,
$$

If $A,B,C$ are $(0,1)$-forms then,  
\beq
A_\mb \,B_\nb\, C_\rb\,\, \O^{\mb\nb\rb} \star 1 ~=~ \ii\, \O\, \,A  B  C~, \qquad A  B  C ~=~ \frac{1}{||\O||^2}\, A_\mb\, B_\nb\, C_\rb \,\,\O^{\mb\nb\rb}\,\, \Ob~.
\label{eq:wedgeRelation}\eeq
where we have the compatibility relation
$$
\frac{\ii \O \, \Ob }{||\O||^2} ~=~ \frac{1}{3!} \,\o^3 ~=~  \vol~.
$$ 
It is also useful to note
 $$
 \starb \O = \ii \O~, \qquad \O \starb \O = \ON^2 \star 1 = \ii \O \Ob~.
 $$
In coordinates
\beq
\star 1 =\frac{\sqrt{g}} {(3!)^2} \ii\e_{\m_1\m_2 \m_3} \e_{\nb_1\nb_2 \nb_3} \dd x^{\m_1} \dd x^{\m_2}\dd x^{\m_3}  \dd x^{\nb_1}\dd x^{\nb_2} \dd x^{\nb_3}.
\label{eq:3foldComplexVol}
\eeq

\newpage
\section{Spinors}
\label{app:Spinors}
We establish some conventions and results for spinors in $d=4$, $d=6$ and $d=10$.

We define the Pauli matrices as
\beq\label{eq:Pauli}
\begin{split}
 \s^1 ~&=~ 
\begin{pmatrix}
 0 & 1\\
 1 & 0
\end{pmatrix},
\qquad 
 \s^2 ~=~ 
\begin{pmatrix}
 0 & -\ii\\
 \ii & 0
\end{pmatrix},
\qquad
  \s^3 ~=~ 
\begin{pmatrix}
 1 & 0\\
 0 & -1
\end{pmatrix}~,
\cr
\end{split}
\eeq
and we denote $\id_n$ the $n\times n$ identity matrix. We also define
$$
\s^0 ~=~ 
\begin{pmatrix}
 -1 & 0 \\
 0 & -1
\end{pmatrix}~.
$$
We denote $\s^{e f} = \half (\s^{e} \s^{f} - \s^{f} \s^{e})$. A similar definition applies for the gamma matrices. 

\subsection{Spinors in flat space}
\subsubsection{$\so(9,1)$}
The Dirac representation of $\so(9,1)$ is 32-dimensional. We denote the 32-dimensional $\so(9,1)$ gamma matrices $\G^M$ and chirality operator $\G = \prod_{M=0}^9 \G^M$. The Dirac spinor decomposes into two Majorana--Weyl representations $\rep{32} = \rep{16}\oplus \rep{16}'$.  Our notation will be that primed representations are negative chirality spinors; unprimed representations are positive chirality spinors. 

Let $\ve$ be Weyl spinor that is of positive chirality $\G\ve = \pm\, \ve$.  As $(\G^M)^*$ and $-(\G^M)^*$ both satisfy the same Lorentz algebra as $\G^M$, there are two  similarity transformations preserving the Lorentz algebra
\beq\label{eq:ComplexConjGamma}
B_{(1)} \G^M B_{(1)}^{-1} = (\G^M)^*~, \qquad B_{(2)} \G^M B_{(2)}^{-1} = -(\G^M)^*~,
\eeq
under which the spinor transforms to $\ve \to B_i \ve$.  Hence,  $\ve^*$ and $B_i \ve$ transform in the same way under Lorentz transformations, and we can define Majorana conjugation to be
$$
\ve^c = B_{(i)}^{-1} \ve^*~,
$$
 and the Majorana condition is  $\ve = \ve^c$.  Applying Majorana conjugation twice gives a consistency condition $B_i^* B_i = 1$, no sum on the $i$, which must be satisfied. For $\so(9,1)$ it is possible to find both $B_{(1)}$ and $B_{(2)}$ satisfying \eqref{eq:ComplexConjGamma} that also satisfy the consistency condition; this is not true for $\so(3,1)$ and $\so(6)$. In the text we utilise $B_{(2)}$, which, with our choice of basis, gives a manifestly consistent Majorana condition for $\so(3,1)\oplus \so(6)$. 
 
We utilise the convention of complex conjugation of a pair of spinors interchanges their order without introducing a sign.

\subsubsection{$\so(3,1)$}
 We work with mostly positive signature. A basis of Dirac gamma matrices are
\beq
\label{eq:so31}
\g^e ~=~ 
\begin{pmatrix}
 0& \s^e\\
 -\sb^e & 0
\end{pmatrix}~,
\eeq
where $\s^e = (\s^0, \s^1, \s^2, \s^3)$ with $\s^0= -\id_2$, and the remaining matrices are the Pauli matrices \eqref{eq:Pauli}. The conjugate matrices $\sb^0 = (\s^0, -\s^1, -\s^2, -\s^3)$. $\g^0$ is anti-hermitian $(\g^0)^2 = -\id_4$ while $\g^1,\cdots,\g^3$ are all hermitian. Complex conjugation is the same as transpose: $(\s^e)^* = (\s^e)^t$. We denote $\g^{ef} = \half (\g^e \g^f - \g^f \g^e)$. 

The chirality matrix is 
$$
\g^{(4)} = -\ii \g^0 \cdots \g^3 ~=~ 
\begin{pmatrix}
 -\id_2 & 0 \\
 0 & \id_2
\end{pmatrix}~.
$$
The Majorana conjugate of a Dirac spinor $\Psi$ is
\beq\label{eq:b4}
\Psi^c ~=~ B_4^{-1} \Psi^*~,\qquad B_4 ~=~ 
\begin{pmatrix}
 0 & - \ve \\
 \ve & 0
\end{pmatrix}~,
\eeq
where $\ve = \ii \s^2$. 
It can be checked that $B_4 \g^e B_4^{-1} = (\g^e)^*$ and that $B_4 B_4^* = 1$ so that $(\Psi^c)^c = \Psi$. 

The  $\rep{4}$ of $\so(3,1)$ admits a pair of Weyl representations
$$
\rep{4} = \rep{2} \oplus \rep{2}'~, \qquad \qquad \Psi ~=~ 
\begin{pmatrix}
 0 \\ \z
\end{pmatrix} 
+
\begin{pmatrix}
 \z' \\ 0
\end{pmatrix}~.
$$
of positive and negative chirality respectively, and in the second equality, expressed as spinors in the basis \eqref{eq:so31}. This is sometimes denoted $\Psi = \z\oplus \z'$.

Where possible, we adopt the 2-component spinor notation, see for example \cite{WessBagger,Argyres:2001eva}. The  indices on Weyl spinors are denoted by $\adot$ and $a$. The rule for raising and lowering is through the $\e$ permutation symbol where $\e^{12} = \e_{21} = 1$ and $\e^{21} = \e_{12} = -1$: 
\beq
\begin{split}
\z'^a ~&=~ \e^{ab} \z'_b~~, \qquad \z'_a = \e_{ab} \z'^b~,\\
 \zeb^\adot ~&=~ \e^{\adot\bdot} \zeb_\bdot~, \qquad \zeb_\adot ~=~ \e_{\adot\bdot} \zeb^\bdot~.
\end{split}
\eeq
Complex conjugation exchanges dots on indices $(\z^a)^* = \zeb^\adot$, $(\e_{ab})^* = \e_{\adot\bdot}$ etc. These spinors are assigned the Grassmann odd property and so anticommute. However, when complex conjugating a pair of spinors, the order is interchanged without a sign: $(\z_a \z'_b)^* =  \zeb'_\bdot \zeb_\adot$. 

The indices on $\s^e$ and $\sb^e$ are related 
$$
\s^e_{a\adot} ~=~ \e_{a b} \sb^{f\,\bdot b} \e_{\adot\bdot}~,
$$ 
and the index structure on $\Psi$  is
\beq
\Psi ~=~ 
\begin{pmatrix}
 \z'_a \\
 \zeb^\adot
\end{pmatrix}~.
\label{eq:Dirac}
\eeq
When indices are not written there is an implicit contraction through the $\e$ symbol. For example
\beq\label{eq:spinorConv}
\z \z' ~=~ \z^a \z'_a = \e^{ab}\z_b \z'_a ~=~ - \e^{ab} \z'_a \z_b ~=~ \z'{}^b \z_b ~=~ \z' \z~.
\eeq
Analogous conventions exist for dotted indices, given by complex conjugating the above equation. Some useful spinor relations to be used inside actions are:
\beq
\label{eq:so31spinorrelations1}
\begin{split}
 \z' \s^e \del_e \zeb ~=~ \zeb \sb^e \del_e \z'~, \qquad (-\ii \z' \s^e \del_e \zeb)^* ~=~ - \ii \z \s^e \del_e \zeb' ~=~ - \ii \zeb' \sb^e \del_e \z~.
\end{split}
\eeq
The Dirac conjugate of a Dirac spinor is
$$
\Psib ~=~ \Psi^\dag \g^0~.
$$
A slight abuse of notation: the bar on a Dirac spinor denotes the Dirac conjugate, while the bar on a Weyl spinor denotes a dotted index. 

The kinetic term for the Dirac spinor in terms of Weyl spinors \eqref{eq:Dirac}
\beq
\label{eq:so31diracKE}
\ii \Psib \g^e \del_e \Psi ~=~ \ii \zeb' \sb^e \del_e \z' + \ii \z \s^e\del_e \zeb'~.
\eeq

A lorentz transformation on a Dirac spinor is $\d \Psi = \L_{ef} \g^{ef}$, with $\L_{ef} = - \L_{fe}$  and in the basis   \eqref{eq:so31} becomes an action on Weyl spinors
\beq
\begin{pmatrix}
 \d \z_a' \\
 \d \zeb^\adot 
\end{pmatrix}
~=~   \L_{ef}
\begin{pmatrix}
 - (\s^{e} \sb^{f})_a{}^b & 0 \\
 0 & - (\sb^{e} \s^{f})^\adot{}_\bdot 
\end{pmatrix} 
\begin{pmatrix}
 \z_b' \\
 \zeb^\bdot 
\end{pmatrix}~,
\label{eq:LorentzWeyl}
\eeq
from which we identify the transformation properties of  $\z_a'$ and $\zeb^\adot$, and identify these with the $\rep{2}'$ and $\rep{2}$ of $\so(3,1)$ respectively. 

The Majorana conjugate is
\beq\label{eq:so31Majorana2}
\Psi ~=~ 
\begin{pmatrix}
 \z'_a \\
 \zeb^\adot
\end{pmatrix}
\qquad\qquad
\Psi^c ~=~ B^{-1} \Psi^* ~=~ 
\begin{pmatrix}
 \z_a\\
 \zeb'^\adot
\end{pmatrix}
\eeq
and as expected $\Psi$ and $\Psi^c$ have the same index structure, with  Majorana conjugation swaping the prime,  reflecting the fact they transform in the same way under Lorentz transformations. Notice that complex conjugation of a Weyl spinor does not by itself give another Weyl spinor. For example,  $\z_a$ transforms as the $\rep{2}'$ but $\zeb_\adot$ does not transform as a $\rep{2}$, as can be seen by conjugating the top line of \eqref{eq:LorentzWeyl} and comparing with the second line. Instead,  one is to complex conjugate and contract with $\e$:   $(\e^{ab} \z_b)^* = \e^{\adot\bdot} \zeb_\bdot$ transforms in the $\rep{2}$.  

In this basis, a Majorana spinor satisfies $\z_a= \z_a'$. We will not impose this and choose to work in the Weyl basis. 

\subsubsection{$\so(6)$}
We describe $\so(6)$ spinors first in flat space, before coupling them to a curved manifold with $\su(3)$--structure in the next section. 

A Dirac spinor decomposes into a pair of Weyl representations
$$
\rep{8} ~=~ \rep{4} \oplus \rep{4}'~.
$$
 A basis compatible with this is
\beq
\begin{split}
\g^1 ~&=~ \s^1 \otimes \s^3 \otimes \s^3~, \quad \g^2 ~=~ \s^2 \otimes \s^3 \otimes \s^3~, \quad \g^3 ~=~ \id_2 \otimes \s^1 \otimes \s^3~, \cr
 \g^4 ~&=~ \id_2\otimes \s^2 \otimes \s^3~,  \, \quad \g^5 ~=~ \id_2 \otimes \id_2 \otimes \s^1~, \quad \g^6 ~=~ \id_2 \otimes \id_2 \otimes \s^2~.\cr\label{eq:so6}
\end{split}
\eeq
The chirality operator is 
\beq
\g^{(6)} ~=~ \ii \g^1 \ldots \g^6 = \s^3 \otimes \s^3 \otimes \s^3~.
\eeq 
We define raising and lowering operators, introducing holomorphic and antiholomorphic indices 
$$
\g^\m = \half (\g^{2\m-1} + \ii \g^{2\m})~, \qquad \g^\mb = \half (\g^{2\mb-1} - \ii \g^{2\mb})~.
$$
These are real $(\g^\m)^* = \g^\m$, related by  $(\g^\m)^\dag = \g^\mb$ and satisfy $\{ \g^\m, \g^\nb\} = \d^{\m\nb}$. In this basis only $\so(2)\oplus \so(2)\oplus \so(2)\subset$ is manifestly preserved. 

The conjugation matrix $B_6$ satisfies
$$
B_6 \g^m B_6^{-1} ~=~ - (\g^m)^*~,
$$
and in this basis is the product of all imaginary matrices so that 
\beq\label{eq:B6so6}
B_6 ~=~ \g^2 \g^4 \g^6 = \s^2 \otimes -\ii \s^1 \otimes \s^2 ~=~ \ii \prod_{\m=1}^3 (\g^\m - \g^\mb)~,
\eeq
which satisfies $ B_6^{-1}  {=} B_6^* {=} -B_6$ and so an $\so(6)$ spinor $\l$ satisfies $(\l^c)^c = \l$. In the second equality, we have written $B_6$ in terms of raising and lowering operators.
The conjugation matrix changes chirality $B_6 \g^\m_\pm {=} \g^\m_\mp B_6$.

\subsection{Spinors on $\IR^{3,1}\times \ccX$}
 In discussing $\IR^{3,1}\times \ccX$, we take the 32-dimensional gamma matrices to decompose 
 $$
 \G^e ~=~ \g^e \otimes \g^6~, \qquad \G^m ~=~ \id \otimes \g^m~, 
 $$
 where $\g^e$ are four-dimensional matrices in \eqref{eq:so31} and $\g^m$ are the 8-dimensional matrices in \eqref{eq:so6}. The chirality matrix is 
 $$
 \G ~=~ \prod_M \G^M ~=~ \g^{(4)} \otimes \g^{(6)}~.
 $$
 The complex conjugation matrix
 \beq
 B ~=~ B_4 \otimes B_6~,
 \eeq
where $B_4$ is in \eqref{eq:b4} and $B_6$ is in \eqref{eq:B6so6}. $B$ is imaginary, unitary and antihermitian $B^\dag = B^{-1} = -B$, and satisfies the property that
 $$
 B \G^M B^{-1} ~=~ - (\G^M)^*~.
 $$
 The Majorana conjugate of a spinor $\ve$ in the $\rep{32}$ is
 $
 \ve^c ~=~ B^{-1} \ve^*~.
 $
 Expand $\ve$ in terms of $\so(3,1)\oplus\so(6)$ spinors
 $$
 \ve ~=~ \z\otimes \l \oplus \z' \otimes \l' ~\cong~ 
\begin{pmatrix}
 0\\
 \zeb^\adot
\end{pmatrix}\otimes \l 
+
\begin{pmatrix}
 \z_a'\\
 0
\end{pmatrix}\otimes \l'~,
 $$
where $\l$ and $\l'$ are  in the $\rep{4}$ and $\rep{4}'$ respectively while $\z$ and $\z'$ are in the $\rep{2}$ and $\rep{2}'$ respectively. In the second line we have written this in terms of the four-dimensional $\so(3,1)$ spinors, with their spinor indices explicit; we will always leave the $\so(6)$ spinor indices implicit.

The Majorana condition $\ve^c = \ve$ is simplified by using \eqref{eq:so31Majorana2}, 
\beq\label{eq:so91Majorana}
\begin{split}
\begin{pmatrix}
 0 \\ 
 \zeb^\adot
\end{pmatrix}
 \otimes \l
 ~=~ 
\begin{pmatrix}
 0 \\
 \zeb'{}^\adot
\end{pmatrix}
\otimes \l'{}^c~,
\qquad
 \begin{pmatrix}
 \z_a' \\ 
 0
\end{pmatrix}
 \otimes \l'
 ~&=~ 
\begin{pmatrix}
 \z_a \\ 
 0
\end{pmatrix}
\otimes \l^c~.
\end{split}
\eeq	
It sometimes more convenient to write this simply as
\beq\label{eq:so91Majorana2}
\zeb^\adot \otimes \l ~=~ \zeb'{}^\adot \otimes \l'^c~.
\eeq
The Majorana-Weyl spinor $\ve$ can now be written solely in terms of say $\z',\l'$:
$$
 \ve ~=~
\begin{pmatrix}
 0\\
 \zeb'^\adot
\end{pmatrix}\otimes \l'{}^c 
+
\begin{pmatrix}
 \z_a'\\
 0
\end{pmatrix}\otimes \l'~,
$$

\subsection{Spinors on a complex manifold $\ccX$ with $\su(3)$--structure}

The manifold $\ccX$ is endowed with an $\su(3)$--structure meaning there is a globally well-defined non-vanishing spinor implying a reduction of the structure group 
$
\so(6) \to \su(3)
$
under which   
\beq\label{eq:Su4branching}
\rep{4} = \rep{3} \oplus \rep{1}~, \qquad \rep{4}' = \brep{3} \oplus \rep{1}~,
\eeq
and the spinors decompose respectively as
$$
\l =  \l_{\rep{3}} \oplus  \l_+~,\qquad\qquad \l' =  \l_{\brep{3}} \oplus  \l_-~.
$$
The spinors $\l_+, \l_-$ are the $\su(3)$ invariant spinors that are nowhere vanishing on the manifold $\ccX$ that define the $\su(3)$--structure. 

With respect to the basis \eqref{eq:so6}, the raising and lowering matrices $\g_+^\m$ and $\g_-^\nb$  are real  and  related by hermitian conjugation  $(\g_+^\m )^\dag = \g_-^\mb$. This reality property is consequence of our choice of basis, and any physical result will not depend on this choice. Care must be taken when interpreting the holomorphy of indices, and where any ambiguity may arise, will keep the $\pm$ subscript. Nonetheless, at the end of a calculation we will be able to interpret the indices in terms of holomorphic or antiholomorphic indices of $\ccT_\ccX^{1,0}\oplus \ccT_\ccX^{0,1}$. 

The matrices satisfy an algebra
$$
\{ \g^\m, \g^\nb \} ~=~ g^{\m\nb}~,
$$
where $\m,\nb$ are coordinate indices and the right hand side is the inverse metric.\footnote{We can phrase this in terms of tangent space indices, and then use the veilbein to goto coordinate indices, but for succinctness have skipped this step.}

Majorana conjugation is defined using a covariant version of $B$ in \eqref{eq:B6so6}. On an $\su(3)$ manifold,  $B$ is a coordinate scalar, gauge invariant, and satisfies the property that
$$
B \g^m B^{-1} ~=~ - (\g^m)^*, \qquad B^* B ~=~ 1~, \quad B^\dag  ~=~ B^{-1}~.
$$
This fixes
\beq
B ~=~ \ii g^{1/4} \left( \frac{1}{3!} \e_{\m\n\r} \g_+^{\m\n\r}  + \frac{1}{2} \e_{\m\n\r} \g_+^\m \g_-^{\n\r} -\frac{1}{2} \e_{\m\n\r}  \g_+^{\m\n} \g_-^\r -  \frac{1}{3!} \e_{{\m\n\r}} \g_-^{{\m\n\r}}\right)
\eeq
This is the main example where confusion can arise in holomorphy of indices, and so we use the $\pm$ subscript for clarity. 

We build spinor representations by lowering and raising operators. Denote the lowest weight state $\l_-$, satisfying $\g^\mb \l_- = 0$. We define the remaining spinors as follows:
\beq\label{eq:internalspinors}
\begin{split}
 \l_-& \qquad\qquad \rep{1}~,\\
 \l_{\rep{3}}:=  \L_\m \g^\m \l_-& \qquad \qquad \rep{3}~,\\
\l_{\brep{3}} ~:=~ \half \L_{\m\n} \g^{\m\n} \l_- &\qquad \qquad \brep{3}~,\\
\l_+:= \frac{1}{3!} \L_+ \e_{\m\n\r}\g^{\m\n\r}  \l_- &\qquad \qquad \rep{1}~.
\end{split}
\eeq 
where $\g^{\m\n} = \half (\g^\m \g^\n - \g^\n\g^\m)$ and $\g^{\m\n\r} = \smallfrac{1}{3!} (\g^\m\g^\n\g^\r - \g^\m \g^\r \g^\n + \cdots)$. Note that $\g^{\m\n} \l_- = \g^\m \g^\n \l_-$. Here $\e_{\m\n\r}$ is the permutation symbol with $\e_{123} = 1$ and $\L_+$ is a tensor density to be fixed. 

To identify $\L_+$ we study its transformation properties under symmetries of the moduli space and under holomorphisms. First note that $\L_+$ transforms like $g^{1/4}$ under holomorphisms. Hence, $\L_+ \propto g^{1/4}$ up to a parameter dependent coordinate scalar.   Second, recall the gauge symmetry $\O \to \m \O$ where $\m = |\m| e^{\ii \x} \in \IC^*$. Under this symmetry the fermions $\l_\pm$ are charged transforming under the $U(1)\subset \IC^*$ as
\footnote{This charge assignment is determined by studying the \K transformations of the \K potential 
$$
\ccK ~=~ -\log\Big(\ii \int \O \,\Ob\Big) - \log \Big(\frac{4}{3} \int \o^3\Big)~.
$$
under $\O \to \mu \O$. 
As described in \cite{WessBagger}, in order to couple $d=4$ chiral fields to gravity preserving $\cN=1$ supersymmetry the $\IR^{3,1}$ fermions must transform, which in order for the $\so(9,1)$ fermions to remain neutral, implies the transformation law \eqref{eq:lpmcharge}.}
\beq
\l_\pm \to \l_\pm e^{\ii \x/2}~,\label{eq:lpmcharge}
\eeq 
and so  $\L_+$ transforms as $\L_+ \to \L_+ e^{\ii \x}$. Hence, $\L_+ \propto (f \sqrt{g}) / |f|$, now fixed up to a gauge--neutral coordinate scalar.  If we demand that $\l_+^\dag \l_+ = \l_-^\dag \l_-$ this fixes the constant to be a phase and we can write the final result as
\beq
\L_+  ~=~ e^{\ii \phi} \frac{f }{||\O||}  ~,  
\eeq
for some phase $e^{\ii \phi}$. There are three phases of interest:  $\psi_\pm = \arg \l_\pm$ and $\zeta$ the phase of $f = |f| e^{\ii \zeta}$. The gauge symmetry eliminates one of these degrees of freedom, and we can form two gauge invariant combinations $\phi = \psi_+ - \psi_- - \zeta$ and $\Psi = \psi_+  + \psi_-$.

Using one of these global symmetries we could choose $\phi = 0$, which in the gauge where $\zeta = 0$ amounts to fixing the relative phases of $\l_\pm$ equal.

%
We state the final result as
\beq
\l_+ ~=~ \frac{e^{\ii\phi} }{||\O||} \frac{1}{3!} \O_{\m\n\r} \g^{\m\n\r} \l_-~,
\eeq
The norms $\l_\pm^\dag \l_\pm$ are gauge invariant coordinate scalars, and so we are free to fix them to be unity
\beq
\l_\pm^\dag \l_\pm ~=~ 1~.
\eeq
We note that $\l_\brep{3}$ can  be written as  
\beq\label{eq:lbrep}
\l_{\brep{3}} ~=~ \L'_\mb \g^\mb \l_+~, \quad \L_\mb' ~=~ \frac{e^{-\ii\phi}}{2||\O||}  \Ob_{\overline{\m\n\r}}\L^{\overline{\n\r}} ~, \quad \L_{\m\n} ~=~ \frac{e^{\ii\phi}}{||\O||} \O_{\m\n}{}^\rb \L_\rb' ~.
\eeq 
By studying $\l_+^\dag \l_+$ in two different ways we identify
\beq
\label{eq:OmegaSpinor}
  \O_{\m\n\r}~=~ - e^{-\ii\phi}\,||\O||\, \l_-^\dag \g_{\m\n\r} \l_+~,\qquad  \Ob_{\overline{\m\n\r}}~=~  e^{\ii\phi}\,||\O||\, \l_+^\dag \g_{\ol{\m\n\r}} \l_-~,
\eeq 
as well as 
\beq
\label{eq:MetricBilinear}
\l_+^\dag \g^\m\g^\nb \l_+ ~=~ g^{\m\nb}~, \qquad\l_-^\dag \g^\nb\g^\m \l_- ~=~ g^{\m\nb}~. 
\eeq

The norms of spinors are then
\beq
\l_{\rep{3}}^\dag \l_{\rep{3}} ~=~ \L_\m \L^\m~, \qquad \l_{\brep{3}}^\dag \l_{\brep{3}} ~=~ \L_\mb' \L^\mb ~=~ \half \L_{\m\n}\L^{\m\n}~.
\eeq

It is useful to tabulate Majorana conjugates of spinors $\l^c = B^{-1} \l^* = (B\l)^*$:
\beq
\label{eq:MajoranaConjugates}
\begin{split}
 \l_-^c ~&=~ -\ii  g^{1/4} \L_+^{*\,-1} \l_+^* ~=~-\ii  \l_+~, \\
   \l_+^c ~&=~ -\ii  g^{-1/4} \L_+^* \l_-^* ~=~ -\ii  \l_-~,\\
\l_{\brep{3}}^c ~&=~ \ii  g^{-1/4}\L_+^* \L'{}_\m^* \g^\m \l_-^* ~=~ \ii \L'{}_\m^* \g^\m \l_-~, \\
\l_{\rep{3}}^c ~&=~ \ii  g^{1/4}\L_+^{*\,-1} \L{}_\mb^* \g^\mb \l_+^* ~=~ \ii  \L_\mb^* \g^\mb \l_+~, 
\end{split}
\eeq
and 
\beq
(\l_{\brep{3}}^c)^{\dag} ~=~ -\ii  \L'{}_\mb \l_-^\dag  \g^\mb~, \qquad (\l_{\rep{3}}^c)^{\dag}  ~=~ - \ii  \L_\m  \l_+^\dag \g^\m~.
\eeq
Finally, given a derivative operator $D_\nb$, which in the text becomes a covariant derivative with respect to the bundle symmetries, we will need the following bilinear 
\beq
\label{eq:yukawabilinear}
(\l_{\brep{3}}^c)^{\dag} \g^\nb D_\nb \l_{\brep{3}} ~=~ \ii e^{\ii\phi}\,\Big( \L_\mb' D_\nb \L_\rb' \Big) \,\frac{ \O^{\ol{\m\n\r}} }{\ON}~, \quad (\l_{\rep{3}}^c)^{\dag} \g^\n D_\n \l_{\rep{3}} ~= -\ii e^{-\ii\phi}\,\Big( \L_\m D_\n \L_\r \Big) \,\frac{ \Ob^{\m\n\r} }{\ON}~.  
\eeq

\subsection{Spinors charged under gauge symmetries}
Sometimes the spinors carry additional structure, for example being charged in a representation of $\Lg \oplus \Lh$.  In that case complex conjugation is promoted to hermitian conjugation. 

Consider first $\z,\z'$; these spinors may be charged in a representation of $\Lg$. In computing a Majorana conjugate, the complex conjugate in \eqref{eq:so31Majorana2} is promoted to hermitian conjugate on the gauge structure. It is normally easiest to do this with the spinor indices explicit; Majorana conjugation does not transpose the spinor structure. 

As a way of illustration, There are three relevant cases to the text. The first are when $\z,\z'$ are singlets under $\ad_\Lg$, in which Majorana conjugation is unchanged from the previous subsection. The second case, $\z$ is in a representation $\rep{R}$, denoted $\z'_{\rep{R'}}$, and hermitian conjugation acts as $(\z_\rep{R}^a)^\dag ~\cong~ \zeb^\adot_{\brep{R}}$.  The third case is when $\z'$ is in the adjoint of $\Lg$, in which it is antihermitian  $(\z_{\ad_\Lg}^a)^\dag =- \zeb^\adot_{\ad_\Lg}$. The Majorana conjugates of the last two cases are  explicitly:
\beq\label{eq:zetaMajConj}
\begin{pmatrix}
\z'_{\rep{R'}\,a} \\
0 
\end{pmatrix}^c ~=~ B_4^{-1} \begin{pmatrix}
(\z'_{\rep{R}'\,a})^\dag \\
0 
\end{pmatrix} ~=~ 
\begin{pmatrix}
 0 \\
 \zeb'{}^\adot_\brep{R'}
\end{pmatrix}~,\qquad
\begin{pmatrix}
\z'_{\ad_\Lg\,a} \\
0 
\end{pmatrix}^c 
~=- 
\begin{pmatrix}
 0 \\
 \zeb'{}^\adot{}
\end{pmatrix}~.
\eeq
The Majorana condition  \eqref{eq:so91Majorana2} implies  that if $\z'$ is in the $\rep{R}$ then $\z$ is in the $\brep{R}$. Of course, if the representation is real, then $\rep{R} = \brep{R}$ in the above.


 Similar comments apply when $\l,\l'$ carry representations of $\Lh$. Only $\l_{\rep{3}}$ and $\l_{\brep{3}}$ turn out to carry non-trivial representations of $\Lh$, and this is through the object $\L_\m$ and $\L'_\mb$ in \eqref{eq:internalspinors} and \eqref{eq:lbrep}.   The singlets $\l_\pm$ are  always gauge singlets. 
The generalisation of \eqref{eq:MajoranaConjugates} is 
\beq
\label{eq:lambdaconj}
\begin{split}
 \l_\pm^c ~&= -\ii  \l_\mp~, \qquad \l_{\brep{3}}^c ~=~  \ii \L'{}_\m^\dag \g^\m \l_-~, \qquad \l_{\rep{3}}^c ~=~ \ii  \L_\mb^\dag \g^\mb \l_+~, 
\end{split}
\eeq
where $\L_\m$ is charged in a representation and $\L_\mb^\dag$ is the appropriate hermitian conjugate. 

Putting this into  \eqref{eq:so91Majorana2} determines $\z\otimes\l$ in terms of $\z'\otimes\l'$:
\beq
\zeb^\adot \otimes \l_+ ~= -\ii \zeb'{}^{\adot\,\dag} \otimes \l_+~, \quad \zeb^\adot \otimes \L_\m \g^\m\l_- ~=~\ii \zeb'{}^{\adot\,\dag} \otimes \L'{}^\dag_\m \g^\m \l_-~.
\eeq


\subsection{Some useful spinor bilinears}

We  express the Majorana--Weyl spinor $\ve$ in terms of $\z',\l'$, and list some bilinears relevant to the main text.
\beq\label{eq:SpinorBilinears1}
\begin{split}
 \ve ~&=~
\begin{pmatrix}
 0\\
 \zeb'^\adot
\end{pmatrix}\otimes \l'{}^c 
+
\begin{pmatrix}
 \z_a'\\
 0
\end{pmatrix}\otimes \l'~,\\
 \veb ~&=~  \begin{pmatrix}
 \z'{}^{\dag\,a} & 0
\end{pmatrix} \otimes (\l'{}^c)^\dag 
+ 
\begin{pmatrix}
 0 & \zeb'^\dag_\adot
\end{pmatrix} \otimes \l'{}^\dag~,\\
\veb \ve &=~ 0
\\
\veb \G^e \del_e \ve   &=~  \Big( \z'{}^{\dag\,a} \,\s_{a\bdot}^e \,\del_e \,\zeb'^\bdot \Big)\, (\l'{}^c)^\dag\l'{}^c
+
\Big( \zeb'^\dag_\adot\, \sb^{e\,\adot b} \,\del_e\, \z'_b \Big)\, \l'{}^\dag \l'~,\\
\veb \G^m \del_m \ve ~&=~ (\z'^{\dag\,a} \z_a')  (\l'^{c\,\dag}  \g^m\del_m \l') + (\zeb'^\dag_\adot \zeb'^{\adot})( \l'^\dag\g^m\del_m\l'^c)~.
\end{split}
\eeq
We have left the four-dimensional spinor indices for clarity, but will now drop them  in spinor contractions, using  the convention \eqref{eq:spinorConv}.

We can now evaluate these relations for  some specific examples relevant to the text when the spinors are charged in representations of $\Lg \oplus \Lh$:
\begin{enumerate}

 \item $\z'\otimes \l' \in (\ad_\Lg, \rep{1})$ of $\Lg \oplus \Lh$.

Use $(\z'{}^a)^\dag = - \zeb'^{\adot}$, its Majorana conjugate \eqref{eq:zetaMajConj}, as well as $\l'=\l_-$ and $\l'^c = -\ii \l_+$: 
 \beq\label{eq:SpinorBilinearAdG}
\begin{split}
 \ve ~&=
\begin{pmatrix}
 \z_a'\\
 0
\end{pmatrix}\otimes \l_-
+
\ii
\begin{pmatrix}
 0\\
 \zeb'^\adot
\end{pmatrix}\otimes  \l_+~,\\
 \veb ~&=~ 
-\begin{pmatrix}
 0 & \zeb'_\adot
\end{pmatrix} \otimes \l_-^\dag
+
\ii\begin{pmatrix}
 \z'{}^{a} & 0
\end{pmatrix} \otimes \l_+^\dag ~,\\
\ii\, \veb \G^e \del_e \ve   &= -2\ii\, \Big( \zeb'\, \sb^{e} \,\del_e\, \z' \Big)\, \l_-^\dag \l_-
~,\\
\ii \veb \G^m \del_m \ve ~&=~0~,
\end{split}
\eeq
where in the last line $\l_+^\dag \g^\m \l_- = 0$. In the third line we understand this will appear integrated and so use integration by parts  $ \zeb'\, \sb^{e} \,\del_e\, \z' = \z' \,\s^e \,\del_e \,\zeb'$. 

\item  $\z'\otimes  \l_{\brep{3}} \in (\rep{R_i}, \brep{r_i})$ of $\Lg \oplus \Lh$.  

In the text this bilinear  the constituents are  a sum over representations, $\ve = \oplus_i \ve_i$ where  $\ve_i\in(\rep{R}_i, \brep{r}_i)$ and the trace projects onto the natural invariants. There are non-zero invariants as the trace derives from the $\ad_{\Le_8}$ which is real. For example, if $(\rep{R}_i,\brep{r_i})$ is complex representation, then the sum $\oplus_i$ contains both $(\rep{R}_i,\brep{r_i})$ and  its conjugate representation $(\brep{R}_i, \rep{r_i})$, with the trace constructing the natural invariant.

\beq\label{eq:SpinorBilinears1}
\begin{split}
 \ve_i ~&=~ 
\begin{pmatrix}
 \z_{\rep{R_i}}'\\
 0
\end{pmatrix}\otimes \L'_{\brep{r}_i\,\mb} \g^\mb \l_+
 +
 \ii \begin{pmatrix}
 0\\
 \zeb_{\brep{R_i}}'
\end{pmatrix}\otimes \L'{}^\dag_{\rep{r}_i\,\m} \g^\m \l_-\\[3pt]
 \veb_j ~&=~
 \begin{pmatrix}
 0 & \zeb'{}_{\brep{R_j}\,}
\end{pmatrix} \otimes ( \L'{}^\dag_{\rep{r}_j\,\m}  \l^\dag_+ \g^\m)
-
\,\ii \begin{pmatrix}
 \z'{}^{}_{\rep{R_j}} & 0
\end{pmatrix} \otimes (\L_{\brep{r}_j\,_\mb}'  \l_-^\dag \g^\mb) ~,\\[3pt]
\ii \tr \Big( \veb_j \G^e \del_e \ve_i\Big)   &=~ 
\ii\tr_\Lg\Big( \,\zeb'_{\brep{R_j}}\, \sb^{e} \,\del_e\, \z'_{\rep{R_i}} \Big)\, \tr_\Lh\Big(\L'^\dag_{\rep{r}_j\,\m} \L'_{\brep{r}_i\,\nb} \Big) g^{\m\nb}
+\\[2pt]
&\qquad\qquad
+\ii\tr_\Lg \Big( \z'_{\rep{R_j}} \,\s^e \,\del_e \,\zeb'_{\brep{R_i}} \Big)\, \tr_\Lh\Big( \L'^\dag_{\rep{r}_i\,\m} \L'_{\brep{r}_j \,\nb} )  \, g^{\m\nb}~,\\[3pt]
\ii \tr\Big(\veb_j \G^m \del_m \ve_i\Big) ~&= - e^{-\ii \phi} \tr_\Lg(\zeb'{}_{\brep{R_j}} \zeb'{}_{\brep{R_i}})   \tr_\Lh \Big( \L'^\dag_{\rep{r_j}\,\m}\del_\n \L'^\dag_{\rep{r}_i\,\r} \Big)\, \frac{\,\,\Ob^{\m\n\r}}{\ON} \\ 
&\qquad - e^{\ii\phi}  \tr_\Lg\Big(\z'_{\rep{R_j}} \z'_{\rep{R_i}}\Big) \tr_\Lh\Big(\L'_{\brep{r}_j\,\mb} \del_\nb\L'_{\brep{r}_i\,\rb} \Big)\, \frac{\,\,\O^{\mb\nb\rb}}{\ON}~. 
\\[3pt]
\ii \tr\Big(\veb_j \G^m [\d A_m, \ve_i]\Big) ~&= - e^{-\ii \phi} \tr_\Lg(\zeb'{}_{\brep{R_j}} \d \Phi^\Xib \zeb'{}_{\brep{R_i}})   \tr_\Lh \Big( \L'^\dag_{\rep{r_j}\,\m} [\d_\Xib \A_\n^\dag , \L'^\dag_{\rep{r}_i\,\r} ]\Big)\, \frac{\,\,\Ob^{\m\n\r}}{\ON} \\ 
&\qquad - e^{\ii\phi}  \tr_\Lg\Big(\z'_{\rep{R_j}} \d \Phi^\Xi \z'_{\rep{R_i}}\Big) \tr_\Lh\Big(\L'_{\brep{r}_j\,\mb} [ \d_\Xi A_\nb,  \L'_{\brep{r}_i\,\rb}] \Big)\, \frac{\,\,\O^{\mb\nb\rb}}{\ON}~.\end{split}
\raisetag{80pt}
\eeq
 $\tr_\Lg$ and $\tr_\Lh$ descend from the trace over $\Le_8$. They are understood to mean to contract the  $\rep{R}$ and $\rep{r}$ indices in the appropriate way in order to get an invariant; if none exists then the trace vanishes. We use \eqref{eq:yukawabilinear} in the last two lines. We have written $\d A_\mb = \d \Phi^\Xi \d_\Xi A_\mb$ to represent a generalised variation of the $\Le_8$ gauge field. In the text this includes moduli and matter fields e.g. $ \d \Phi^\Xi \d_\Xi A_\mb = Y^\a \cD_\a \A_\mb + C^\x \phi_{\x\,\mb} + D^\t \psi_{\t\,\mb}$. 

\end{enumerate}

\newpage
\section{Some representation theory}
\label{app:Representations}
Consider some  Lie algebra $\Le$ and for this subsection only denote  $a=1, \ldots,\dim \Le$.  
We choose $A$ to be antihermitian $A^\dag = -A$. In terms of adjoint generators $T^a$:
$$
A = A^a T^a, \qquad (A^a)^*  = A^a, \qquad [T^a]^\dag = - T^a.
$$
Anti-hermitian matrices  for the fundamental and anti-fundamental satisfy
$$
[T^a,T^b]=f^{abc}T^c, \quad [(T^*)^a,(T^*)^b]=f^{abc}(T^*)^c,\quad (f^{abc})^* = f^{abc}.
$$
We identify
$$
 T_{\overline R} = (T_R^a)^*.
$$
Let $\chi$ be in the fundamental and $\Psi$ in the anti-fundamental. The covariant derivatives are 
$$
\dd_A \chi = d \chi + A \chi, \qquad \dd_A \psi^t  = d\psi^t + \psi^t A.
$$


 Decompose into type:
$$
A = A_{0,1} + A_{1,0} = (A_{0,1}^a + A_{1,0}^a) T^a.
$$
For $A^a$ to be a real form, we require
\beq
(A^a_{0,1})^* = A_{1,0}^a, \qquad (A^a_{1,0})^* = A_{0,1}^a.
\eeq
This is consistent with $A_{0,1}^\dag = - A_{1,0}$ and $A_{1,0}^\dag = - A_{0,1}$. 

As in the paper we define
$$
A = \A - \A^\dag,
$$
where
$$
\A := A_{0,1}, \qquad \A^\dag := - A_{1,0}.
$$
Now define the components of $\A$ and $\A^\dag$:
\beq
\A = \A^aT^a, \qquad \A^\dag = (\A^\dag)^a T^a. 
\eeq
The conjugation property of $\A^a$ is:
\beq
\begin{split}
 (\A^a)^* = (A_{0,1}^a)^* = A_{1,0}^a = - (\A^\dag)^a.
\end{split}
\eeq
\newpage

\raggedright
\baselineskip=10pt

\providecommand{\href}[2]{#2}\begingroup\raggedright\endgroup

\end{document}